\documentclass[]{fac}

\usepackage{amsmath}
\usepackage{amsfonts}
\usepackage{amssymb}
\usepackage{booktabs}
\usepackage{color}
\usepackage{graphicx}
\usepackage{xspace}
\PassOptionsToPackage{hyphens}{url}\usepackage{hyperref}
\hypersetup{breaklinks=true}
\usepackage{mathtools}
\usepackage{phaistos}
\usepackage{siunitx}
\usepackage{microtype}
\usepackage{wasysym}
\usepackage{subcaption}
\usepackage{pdfpages}

\usepackage{tikz}
\usetikzlibrary{calc}
\usetikzlibrary{fit}
\usetikzlibrary{calc}
\usetikzlibrary{shapes}
\usetikzlibrary{shapes.geometric}
\usetikzlibrary{shapes.gates.logic.US}
\usetikzlibrary{arrows}
\usetikzlibrary{arrows.meta}

\tikzset{
  big edge/.style={
    green,
    thick,
  },
  big edgep/.style={
    big edge,
    -{Circle[fill=black,black,width=2,length=2,sep=-1]}
  },
  big pedge/.style={
    big edge,
    {Circle[fill=black,black,width=2,length=2,sep=-1]}-
  },
  big pedgep/.style={
    big edge,
    {Circle[fill=black,black,width=2,length=2,sep=-1]}-{Circle[fill=black,black,width=2,length=2,sep=-1]}
  },
  big edgec/.style={
    big edge,
    -{Bar[fill=green,green,width=4,length=0,sep=0]}
  },
  big pedgec/.style={
    big edge,
    {Circle[fill=black,black,width=2,length=2,sep=-1]}-{Bar[fill=black,black,width=4,length=0,sep=0]}
  },
  big region/.style={
    draw,
    rectangle,
    rounded corners=1.5,
    dashed,
    dash pattern=on 1pt off 1pt,
    thin,
    gray,
  },
  big site/.style={
    big region,
    fill=gray!60,
    text=black,
  },
  big react/.style={
    black,
    thick,
    -stealth,
    line width=3,
    shorten <=3,
    shorten >=3,
  },
  big react rev/.style={
    black,
    thick,
    stealth-stealth,
    line width=3,
    shorten <=3,
    shorten >=3,
  },
  lbl/.style={
    font=\tiny\sf,
    inner sep=1,
  },
  lbl conc/.style={
    font=\tiny,
    inner sep=1,
  }
}
 
\pgfdeclarelayer{foreground}
\pgfsetlayers{main,foreground}

\tikzset{
 dist/.style={
    draw,
 },
 buffer/.style={
    draw,
  },
 node/.style={
    draw,
    circle,
  },
  sink/.style={
    draw,
    diamond
  },
}

\usepackage{pgfplots}
\usepackage{pgfplotstable}
\usepgfplotslibrary{colorbrewer}
\usetikzlibrary{pgfplots.colorbrewer}

\usepackage{xspace}
\newcommand{\ie}{\emph{i.e.}\@\xspace}
\newcommand{\eg}{\emph{e.g.}\@\xspace}
\makeatletter
\newcommand*{\etc}{\@ifnextchar{.}{etc}{etc.\@\xspace}}
\makeatother
\newcommand{\etal}{\emph{et al.}\@\xspace}

\usepackage[capitalise, nameinlink, noabbrev, sort]{cleveref}

\crefname{lstlisting}{Listing}{Listings}
\Crefname{lstlisting}{Listing}{Listings}

\usepackage{listings}

\definecolor{solarized@base03}{HTML}{002B36}
\definecolor{solarized@base02}{HTML}{073642}
\definecolor{solarized@base01}{HTML}{586e75}
\definecolor{solarized@base00}{HTML}{657b83}
\definecolor{solarized@base0}{HTML}{839496}
\definecolor{solarized@base1}{HTML}{93a1a1}
\definecolor{solarized@base2}{HTML}{EEE8D5}
\definecolor{solarized@base3}{HTML}{FDF6E3}
\definecolor{solarized@yellow}{HTML}{B58900}
\definecolor{solarized@orange}{HTML}{CB4B16}
\definecolor{solarized@red}{HTML}{DC322F}
\definecolor{solarized@magenta}{HTML}{D33682}
\definecolor{solarized@violet}{HTML}{6C71C4}
\definecolor{solarized@blue}{HTML}{268BD2}
\definecolor{solarized@cyan}{HTML}{2AA198}
\definecolor{solarized@green}{HTML}{859900}

\lstnewenvironment{bigrapher}[1][] {
  \lstset{
    basicstyle=\scriptsize\ttfamily,
    mathescape=true,
    breaklines=true,
    sensitive=true,
    numbers=left,
    numberstyle=\tiny,
    escapeinside={@}{@},
    numberstyle=\tiny\color{solarized@base01},
    keywordstyle=\color{solarized@green},
    stringstyle=\color{solarized@cyan}\ttfamily,
    commentstyle=\color{solarized@base01},
    emphstyle=\color{solarized@red},
    showstringspaces=false,
    flexiblecolumns=false,
    frame=tb,
    keywords={fun, brs, end, sbrs, pbrs, nbrs, begin, init, atomic, preds,
      rules, action, big, ctrl, float, int, react},
    morekeywords={in, if, param, ctx},
    comment=[l]{\#},
    #1
  }} {}

\usepackage{tikz}
\usetikzlibrary{shapes}
\usetikzlibrary{arrows}
\usetikzlibrary{fit}
\usetikzlibrary{calc}

\definecolor{safelightblue}{rgb}{0.65098, 0.807843, 0.890196}
\definecolor{safedarkblue}{rgb}{0.121569, 0.470588, 0.705882}
\definecolor{safeverylightblue}{rgb}{0.857843, 0.931961, 0.996078}
\definecolor{safeverydarkblue}{rgb}{0.007843, 0.219608, 0.345098}
\definecolor{safelightorange}{rgb}{0.996078, 0.931961, 0.857843}
\definecolor{safelightgreen}{rgb}{0.698039, 0.87451, 0.541176}
\definecolor{safemediumorange}{rgb}{0.74902, 0.505882, 0.490196}
\definecolor{safelightgreen}{rgb}{0.698039, 0.87451, 0.541176}
\definecolor{safedarkgreen}{rgb}{0.2, 0.627451, 0.172549}
\definecolor{safepurple}{rgb}{0.458824, 0.439216, 0.701961}
\definecolor{safedarkorange}{rgb}{0.34902, 0.176471, 0.015686}
\definecolor{safelightgrey}{rgb}{0.7, 0.7, 0.7}
\definecolor{safenearlywhite}{rgb}{0.9, 0.9, 0.9}
\definecolor{safereallynearlywhite}{rgb}{0.93, 0.93, 0.93}

\definecolor{englishvelvet}{HTML}{58355E}
\definecolor{vermilion}{HTML}{E03516}
\definecolor{canary}{HTML}{FFF689}
\definecolor{teagreen}{HTML}{CFFFB0}
\definecolor{ceruleanfrost}{HTML}{5998C5}
\definecolor{cameopink}{HTML}{DAB6c4}

\newcommand\id{\mathsf{id}}

\newcommand{\conc}[1]{\widetilde{#1}}

\newcommand{\rroccurr}[3]{\sigma_{#3}[#1,#2]}
\newcommand{\rroccur}[2]{\rroccurr{#1}{#2}{\mathsf{R}}}

\newcommand{\totw}[2]{\omega[#1,#2]}
\newcommand{\totwr}[3]{\omega[#1,#2]\!\restriction_{#3}}
\newcommand{\distr}[3]{\mu_{#1}(#2)\!\restriction_{#3}}
\newcommand{\distrf}[2]{\mu_{#1}\!\restriction_{#2}}

\newcommand{\wfail}[0]{w_{f}}
\newcommand{\wcon}[0]{w_{r}}
\newcommand{\wsuc}[0]{w_{s}}

\DeclareMathOperator{\react}{\mathrel{\frac{\raisebox{0.75mm}{\begin{scriptsize}\ensuremath{\hspace*{1mm}\ \hspace*{1mm}}\end{scriptsize}}}{}} \joinrel{\!\!\vartriangleright}}
\newcommand{\reactp}[1]{\operatorname{\mathrel{\frac{\raisebox{0.75mm}{\begin{scriptsize}\ensuremath{\hspace*{1mm}\ #1 \hspace*{1mm}}\end{scriptsize}}}{}} \joinrel{\!\!\vartriangleright}}}
\DeclareMathOperator{\rrul}{\mathrel{\frac{\raisebox{0.75mm}{\begin{scriptsize}\ensuremath{\hspace*{1mm}\ \hspace*{1mm}}\end{scriptsize}}}{}} \joinrel{\!\!\blacktriangleright}}
\newcommand{\rrulp}[1]{\operatorname{\mathrel{\frac{\raisebox{0.75mm}{\begin{scriptsize}\ensuremath{\hspace*{1mm}\ #1 \hspace*{1mm}}\end{scriptsize}}}{}} \joinrel{\!\!\blacktriangleright}}}

\usepackage[disable]{todonotes}

\newcommand{\mscom}[1]{\todo[inline, color=green!40]{MS: #1}}

\newtheorem{definition}{Definition}

\newtheorem{lemma}{Lemma}

\begin{document}

\title{Probabilistic Bigraphs}

\journal{}

\author[Archibald, Calder, and Sevegnani]
{Blair Archibald$^{1}$, Muffy Calder$^{1}$, and  Michele Sevegnani$^{1}$\\
  $^{1}$School of Computing Science, University of Glasgow, UK}

\correspond{Blair Archibald, email: blair.archibald@glasgow.ac.uk}

\makecorrespond \maketitle

\begin{abstract}
  Bigraphs are a universal computational modelling formalism for the spatial and
  temporal evolution of a system in which entities can be added and removed. We
  extend bigraphs to probablistic bigraphs, and then again to action bigraphs,
  which include non-determinism and rewards. The extensions are implemented in
  the BigraphER toolkit and illustrated through examples of virus spread in
  computer networks and data harvesting in wireless sensor systems. BigraphER
  also supports the existing \emph{stochastic bigraphs} extension of Krivine
  \etal, and using BigraphER we give, for the first time, a direct
  implementation of the membrane budding model used to motivate stochastic
  bigraphs \cite{DBLP:journals/entcs/KrivineMT08}.

\end{abstract}

\begin{keywords}
  Bigraphs,
  Probabilistic bigraphical reactive systems,
  Discrete Time Markov Chains,
  Markov Decision Processes
\end{keywords}

\section{Introduction}
\label{sec:intro}

Bigraphical reactive systems (BRSs)~\cite{DBLP:books/daglib/0022395} are a
universal   computational modelling formalism for systems that evolve in
time and space. They consist of bigraphs, a graph based formalism that
models entity relationships, both spatially and through (global) links, and    a rewriting framework that allows models
to evolve over time via a set of reaction (rewrite) rules. Applying a
reaction rule, $L \rrul R$, replaces an occurrence of bigraph $L$ (in a bigraph)  with bigraph $R$. BRSs can represent a diverse range of phenomena including
  mixed-reality games~\cite{DBLP:journals/tochi/BenfordCRS16},
network management~\cite{DBLP:journals/scp/CalderKSS14}, wireless communication
protocols~\cite{DBLP:journals/fac/CalderS14}, biological
processes~\cite{DBLP:journals/entcs/KrivineMT08}, cyber-physical security~\cite{oro47351}, and indoor 
environments~\cite{DBLP:conf/giscience/WaltonW12}.

In practice, the  systems we wish to model may be probabilistic, stochastic, or explicitly make 
non-deterministic choices. Standard BRSs have no notion of the first two  concepts, and are implicitly non-deterministic in that if there is a match to  $L$ then
any  rule can be applied.

Previously, Krivine \etal~\cite{DBLP:journals/entcs/KrivineMT08} extended bigraphs to {\em stochastic bigraphs}, by associating rates (rather than weights) with reaction rules.  We build on  that work,  utilising similar ideas to create
\emph{probabilistic bigraphs} -- a discrete variant. We then  take the theory further to
 allow explicit non-determinism with \emph{action bigraphs} that
encode Markov decision processes~\cite{bellman1957markovian}, by adding
\emph{actions} and \emph{rewards}. 

For each of the three types of system  -- probabilistic, stochastic, and action based --  we
provide an implementation of the theory in the BigraphER toolkit~\cite{DBLP:conf/cav/SevegnaniC16}. This allows, for the first time, an implementation and 
analysis of the Krivine et al stochastic bigraph example~\cite{DBLP:journals/entcs/KrivineMT08} without requiring a separate
PRISM~\cite{kwiatkowska.ea_ProbMobAmbs:2009} model.

We make the following contributions:
\begin{itemize}
  \item we extend standard BRSs with probabilistic reaction rules, to create 
    probabilistic BRSs, 
    
  \item we extend  probabilistic BRSs with non-deterministic actions and reward structures, to  create action BRSs,  
\item we provide an implementation  in
    BigraphER~\cite{DBLP:conf/cav/SevegnaniC16}  of probabilistic, stochastic,
    and action BRSs,
  \item we illustrate the new modelling capability through  examples of  virus spread through computer networks, the membrane
    budding example of~\cite{DBLP:journals/entcs/KrivineMT08}, and data harvesting in wireless sensor networks with mobile sinks.
\end{itemize}

\paragraph{\textbf{Outline}}

The paper is structured as follows. Bigraphs and BRSs are   introduced in
\cref{sec:bigraphs,sec:brss}, with emphasis on the important notion of
matching/occurrence;   probabilistic systems are introduced in
\cref{sec:probsystems}. In \cref{sec:pbrs} we introduce probabilistic BRSs by
adding \emph{relative weights} to reaction rules. \Cref{sec:mdps} extends
probabilistic bigraphs further by adding explicit \emph{actions} that represent
non-deterministic choice. We evaluate the approaches through a set of further  examples,  implemented in an extended BigraphER,
in \cref{sec:examples}. We conclude  in \cref{sec:discussion}  with a discussion
of the limitations of the approaches, how they relate to other probabilistic
modelling frameworks, and future work.

\section{Background}

\subsection{Bigraphs}
\label{sec:bigraphs}

We introduce bigraphs by example,
formal definitions  can be  found
elsewhere~\cite{DBLP:books/daglib/0022395}. Although we restrict ourselves to
Milner's original formulation of bigraphs (standard bigraphs), the
probabilistic, stochastic, and non-deterministic variants are also applicable
to, and implemented for, bigraphs with
sharing~\cite{dblp:journals/tcs/sevegnanic15} -- an extension supporting
overlapping spatial regions.

Bigraphs are a universal computational model for representing both the spatial
configuration of entities, and their non-spatial interactions. A bigraph
consists of two orthogonal structures: the \emph{place graph}, that represents
topological space in terms of containment, and the \emph{link graph}, a
hypergraph that expresses non-spatial relationships among entities. Each entity
has a \emph{type} that determines its (fixed) \emph{arity}, \ie number of links,
and whether it is \emph{atomic}, \ie if it cannot contain other nodes.

\begin{figure}
\centering
    \begin{subfigure}[b]{0.45\textwidth}
    \centering
    \begin{tikzpicture}
      \node[big site] (s1) {};
      \node[circle, draw, above right=0.25 of s1] (a1) {};
      \node[lbl, anchor = south east] at (a1.north west) (a1_lbl) {A};
      \node[draw, ellipse, fit=(s1)(a1)(a1_lbl)] (b1) {};
      \node[lbl, anchor = south east] at (b1.north west) (b1_lbl) {B};
      \node[big region, fit=(b1)(b1_lbl)] (r1) {};

      \node[circle, draw, right=1.55 of s1, yshift=1] (a2) {};
      \node[lbl, anchor = south east] at (a2.north west) (a2_lbl) {A};
      \node[circle, draw, above right=0.2 of a2] (a3) {};
      \node[lbl, anchor = south east] at (a3.north west) (a3_lbl) {A};
      \node[draw, ellipse, fit=(a2)(a3)(a2_lbl)(a3_lbl)] (b2) {};
      \node[lbl, anchor = south west] at (b2.north east) (b2_lbl) {B};
      \node[big region, fit=(b2)(b2_lbl)] (r2) {};

      \node[above=0.2 of b1] (x) {$x$};
      \node[above=0.2 of b2] (y) {$y$};
      \node[below=0.2 of b1] (z) {$z$};

\coordinate (h1) at ($($(a1)!0.5!(a2)$) + (0,0)$);
      \draw[big edge] (a1.east) to[out=0, in=180] (h1) to[out=0, in=180] (a2.west);
      \draw[big edge] (h1) to[in=90, out=-90] (s1) to[out=-90, in=90] (z);

      \draw[big edgec] (a3.west) to[out=180, in=-60] ($(a3.west) + (-0.2,0.1)$);

      \draw[big edge] (b1.north east) to[out=60, in=-90] (x);
      \draw[big edge] (b2.north west) to[out=120, in=-90] (y);
    \end{tikzpicture}
    \caption{$B : \langle 1, \{z\}\rangle \rightarrow \langle 2, \{x, y\}\rangle$}
    \label{fig:bg-example-a}
\end{subfigure}
\begin{subfigure}[b]{0.45\textwidth}
    \centering
    \begin{tikzpicture}
      \node[big site] (s1) {};
      \node[circle, draw, above right=0.25 of s1] (a1) {};
      \node[lbl, anchor = south east] at (a1.north west) (a1_lbl) {A};
      \node[draw, ellipse, fit=(s1)(a1)(a1_lbl)] (b1) {};
      \node[lbl, anchor = south east] at (b1.north west) (b1_lbl) {B};
      \node[big region, fit=(b1)(b1_lbl)] (r1) {};

      \node[above=0.2 of b1, xshift=-10] (x) {$x$};
      \node[above=0.2 of b1, xshift=10] (y) {$y$};
\node[below=0.2 of b1, opacity=0] (z) {$z$};

      \draw[big edge] (b1.north west) to[out=120, in=-90] (x);
      \draw[big edge] (a1.north east) to[out=60, in=-90] (y);
    \end{tikzpicture}
    \caption {$G : \langle 1, \emptyset\rangle \rightarrow \langle 1, \{x, y\}\rangle$}
    \label{fig:bg-example-b}
    \end{subfigure}
    \begin{subfigure}[b]{\textwidth}
  \centering
  \begin{tikzpicture}
    \begin{scope}[every node/.style={opacity=0.3}, every path/.style={opacity=0.3}]
      \node[big site] (s1) {};
      \node[circle, draw, above right=0.25 of s1, opacity=1] (a1) {};
      \node[lbl, anchor = south east, opacity=1] at (a1.north west) (a1_lbl) {A};
      \node[draw, ellipse, fit=(s1)(a1)(a1_lbl), opacity=1] (b1) {};
      \node[lbl, anchor = south east, opacity=1] at (b1.north west) (b1_lbl) {B};
      \node[big region, fit=(b1)(b1_lbl), opacity=1] (r1) {};

      \node[circle, draw, right=1.55 of s1, yshift=1] (a2) {};
      \node[lbl, anchor = south east] at (a2.north west) (a2_lbl) {A};
      \node[circle, draw, above right=0.2 of a2] (a3) {};
      \node[lbl, anchor = south east] at (a3.north west) (a3_lbl) {A};
      \node[draw, ellipse, fit=(a2)(a3)(a2_lbl)(a3_lbl)] (b2) {};
      \node[lbl, anchor = south west] at (b2.north east) (b2_lbl) {B};
      \node[big region, fit=(b2)(b2_lbl)] (r2) {};

      \node[above=0.2 of b1, opacity=1] (x) {$x$};
      \node[above=0.2 of b2] (y) {$y$};
      \node[below=0.2 of b1] (z) {$z$};

\coordinate (h1) at ($($(a1)!0.5!(a2)$) + (0,0)$);
      \node[opacity=1] at (h1) (ym) {$y$};
      \draw[big edge, opacity=1] (a1.east) to[out=0, in=180] (h1);
      \draw[big edge] (h1) to[out=0, in=180] (a2.west);
      \draw[big edge] (h1) to[in=90, out=-90] (s1) to[out=-90, in=90] (z);

      \draw[big edgec] (a3.west) to[out=180, in=-60] ($(a3.west) + (-0.2,0.1)$);

      \draw[big edge, opacity=1] (b1.north west) to[out=120, in=-90] (x);
      \draw[big edge] (b2.north east) to[out=60, in=-90] (y);
    \end{scope}

    \begin{scope}[shift={(4,0)}, every node/.style={opacity=0.3}, every path/.style={opacity=0.3}]
      \node[big site] (s1) {};
      \node[circle, draw, above right=0.25 of s1] (a1) {};
      \node[lbl, anchor = south east] at (a1.north west) (a1_lbl) {A};
      \node[draw, ellipse, fit=(s1)(a1)(a1_lbl)] (b1) {};
      \node[lbl, anchor = south east] at (b1.north west) (b1_lbl) {B};
      \node[big region, fit=(b1)(b1_lbl)] (r1) {};

      \node[circle, draw, right=1.55 of s1, yshift=1, opacity=1] (a2) {};
      \node[lbl, anchor = south east, opacity=1] at (a2.north west) (a2_lbl) {A};
      \node[circle, draw, above right=0.2 of a2] (a3) {};
      \node[lbl, anchor = south east] at (a3.north west) (a3_lbl) {A};
      \node[draw, ellipse, fit=(a2)(a3)(a2_lbl)(a3_lbl), opacity=1] (b2) {};
      \node[lbl, anchor = south west, opacity=1] at (b2.north east) (b2_lbl) {B};
      \node[big region, fit=(b2)(b2_lbl), opacity=1] (r2) {};

      \node[above=0.2 of b1] (x) {$x$};
      \node[above=0.2 of b2, opacity=1] (y) {$x$};
      \node[below=0.2 of b1] (z) {$z$};

\coordinate (h1) at ($($(a1)!0.5!(a2)$) + (0,0)$);
      \node[opacity=1] at (h1) (ym) {$y$};
      \draw[big edge] (a1.east) to[out=0, in=180] (h1);
      \draw[big edge, opacity=1] (h1) to[out=0, in=180] (a2.west);
      \draw[big edge] (h1) to[in=90, out=-90] (s1) to[out=-90, in=90] (z);

      \draw[big edgec] (a3.west) to[out=180, in=-60] ($(a3.west) + (-0.2,0.1)$);

      \draw[big edge] (b1.north west) to[out=120, in=-90] (x);
      \draw[big edge, opacity=1] (b2.north east) to[out=60, in=-90] (y);
    \end{scope}

    \begin{scope}[shift={(8,0)}, every node/.style={opacity=0.3}, every path/.style={opacity=0.3}]
      \node[big site] (s1) {};
      \node[circle, draw, above right=0.25 of s1] (a1) {};
      \node[lbl, anchor = south east] at (a1.north west) (a1_lbl) {A};
      \node[draw, ellipse, fit=(s1)(a1)(a1_lbl)] (b1) {};
      \node[lbl, anchor = south east] at (b1.north west) (b1_lbl) {B};
      \node[big region, fit=(b1)(b1_lbl)] (r1) {};

      \node[circle, draw, right=1.55 of s1, yshift=1] (a2) {};
      \node[lbl, anchor = south east] at (a2.north west) (a2_lbl) {A};
      \node[circle, draw, above right=0.2 of a2, opacity=1] (a3) {};
      \node[lbl, anchor = south east, opacity=1] at (a3.north west) (a3_lbl) {A};
      \node[draw, ellipse, fit=(a2)(a3)(a2_lbl)(a3_lbl), opacity=1] (b2) {};
      \node[lbl, anchor = south west, opacity=1] at (b2.north east) (b2_lbl) {B};
      \node[big region, fit=(b2)(b2_lbl), opacity=1] (r2) {};

      \node[above=0.2 of b1] (x) {$x$};
      \node[above=0.2 of b2, opacity=1] (y) {$x$};
      \node[below=0.2 of b1] (z) {$z$};

\coordinate (h1) at ($($(a1)!0.5!(a2)$) + (0,0)$);
      \draw[big edge] (a1.east) to[out=0, in=180] (h1);
      \draw[big edge] (h1) to[out=0, in=180] (a2.west);
      \draw[big edge] (h1) to[in=90, out=-90] (s1) to[out=-90, in=90] (z);

      \draw[big edgec, opacity=1] (a3.west) to[out=180, in=-60] ($(a3.west) + (-0.2,0.1)$);
      \node[opacity=1] at ($(a3.west) + (-0.32,0.1)$) (ym) {$y$};

      \draw[big edge] (b1.north west) to[out=120, in=-90] (x);
      \draw[big edge, opacity=1] (b2.north east) to[out=60, in=-90] (y);
    \end{scope}
  \end{tikzpicture}

  \caption{Occurrences of $G$ in $B$.
    Note that it is possible to rename links as required.}
  \label{fig:bg-example-matches}
    \end{subfigure}
\caption {Example bigraph $B$ (a), pattern bigraph $G$ (b), and occurrences (c).}
\end{figure}
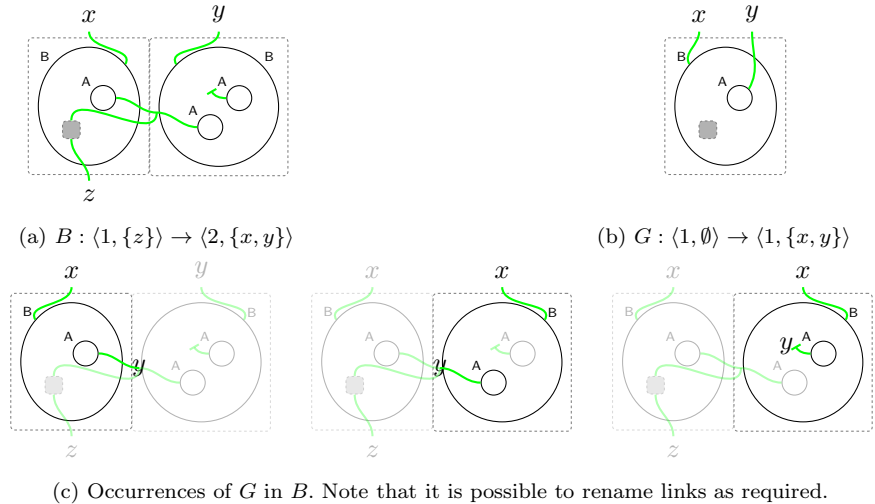

Bigraphs have an equivalent diagrammatic and algebraic notation. Throughout this
paper we use the intuitive diagrammatic notation where possible. An example
bigraph is shown in \cref{fig:bg-example-a}. Entities are drawn as (coloured)
shapes, with the label, \eg \textsf{A}, \textsf{B}, \dots determining the type.
Where it is clear from the context we will often omit the labels. Entities may
be nested, \eg \textsf{A} is inside \textsf{B}, and non-atomic entities can have
any (finite) number of children. The green hyperlinks represent non-spatial
links between entities, such as between the two \textsf{A}'s in different
\textsf{B}'s. Entities have fixed arity, so  the rightmost \textsf{A} in \cref{fig:bg-example-a} \emph{must}
have a single link, but in this case it is \emph{closed}.

Bigraphs are compositional in nature, that is, we may combine smaller bigraphs
to create larger models. To achieve this compositionality, alongside entities,
bigraphs may contain \emph{regions}, shown by clear dashed rectangles, which
represent adjacent parts of a system; \emph{sites}, shown by filled dashed
rectangles,   represent abstraction, \ie an unspecified bigraph (possibly the empty bigraph) exists
there; a set of \emph{inner names}, \eg $\{z\}$, allows names to be connected from
below; and a set of \emph{outer names}, \eg $\{x, y\}$, allows these links to connect
with a wider context. 
Capabilities to interact with an external environment are recorded formally 
in the \emph{interface} of a bigraph. For example, in \cref{fig:bg-example-a} we write 
$B : \langle 1, \{z\}\rangle \rightarrow \langle 2, \{x, y\}\rangle$ 
to indicate that $B$ has one site and inner name $z$ (written $\langle 1, \{z\}\rangle$) and, two regions and outer names $x, y$ (written $ \langle 2, \{x, y\}\rangle$).
We use $X,Y$ and $K, I,J$ to denote sets of names and interfaces, respectively.

Composition of two
bigraphs $F: K \rightarrow I$ and $G: I \rightarrow J$ is written
$$G \circ F: K \rightarrow J$$
and operates by placing the regions on $F$ inside the sites of $G$ and linking like outer names
from $F$ with inner names from $G$. When the name sets are disjoint, bigraphs may also be
combined horizontally by placing regions, which
may contain any other bigraph, side-by-side. This is denoted by 
$$F_0 \otimes F_1:\langle m_0 + m_1, X_0 \cup X_1\rangle \rightarrow \langle n_0 + n_1, Y_0 \cup Y_1\rangle$$
with $F_i:\langle m_i, X_i\rangle \rightarrow \langle n_i, Y_i\rangle$ and $i=0,1$.

We write $\id_X$ for the \emph{identity} bigraph that maps like-names to like-names. 
We call bigraphs with no sites or
inner names, \eg those that cannot be composed with others, \emph{ground}. In general,
we define reactive systems over ground bigraphs, since these represent fully formed
models. We use lowercase letters $f, g, g_{0}, \dots$ to denote ground bigraphs
and uppercase for arbitrary bigraphs (including those that may be ground).

When constructing bigraph models we use \emph{abstract bigraphs}, where
entities are identified using their types, \eg an entity \textsf{A}.  However,
when rewriting models, to identity specific entities, we work instead with \emph{concrete bigraphs},
$\conc{F}, \conc{G}, \dots$, $\conc{g}$, where
entities and closed links have distinct identifiers, \eg $v,u,e,\ldots$. For a
bigraph $G$, we assign an arbitrary \emph{concretion} $\conc{G}$ by giving distinct
labels to entities/closed links. We say two concrete bigraphs $\conc{F}$ and $\conc{G}$ are support
equivalent, denoted $\conc{F} \bumpeq \conc{G}$, if they are equal under a renaming of
entities and links. A bigraph is trivially support equivalent to itself.
An abstract bigraph $B$ is a $\bumpeq$-equivalence class of concrete bigraphs\footnote{In \cite{DBLP:books/daglib/0022395}, lean-support equivalence ($\Bumpeq$) is used instead. It corresponds to support equivalence ($\bumpeq$) after discarding idle links, \ie links connecting zero entities or names.}, $B = [\conc{B}]_{\bumpeq}$ with $\conc{B}$ an arbitrary concretion of $B$.
An example of support equivalence is in \cref{sec:pbrs}.

\subsection{Bigraphical Reactive Systems}
\label{sec:brss}

A bigraph represents a system at a single point in time. To encode the dynamics of a
system we create a Bigraphical Reactive System (BRS) using a set of
reaction rules of the form $L \rrul R$, where $L$ and $R$ are
bigraphs. Intuitively, a BRS operates by finding occurrences of $L$ within a
larger model and replacing these with $R$.

To determine if a bigraph $L$ is present in a bigraph $B$ we need the
following definition, which applies to both concrete and abstract bigraphs.

\begin{definition}[occurrence]\label{def:occurrence}
  A bigraph $L$ occurs in bigraph $B$ if the
  equation $B = C \circ (L \otimes \id_{X}) \circ D$ holds for some
  set of names $X$ and bigraphs $C$ and $D$. Two occurrences are equal if
  they differ only by a permutation or a bijective renaming on the
  composition interfaces; otherwise they are distinct.
\end{definition}

The use of the identity bigraph $\id_{X}$ allows links to pass between $C$ and $D$. An important property is that it is possible to determine an abstract
occurrence starting from a concrete one. In other words, a bigraph $L$
occurs in $B$ only if an arbitrary concretion $\conc{L}$ occurs in an
arbitrary concretion $\conc{B}$. In a given bigraph, there may be multiple
occurrences of another bigraph. 
For example, in \cref{fig:bg-example-matches}, bigraph $G$
occurs three times within bigraph $B$.

In general, the decomposition corresponding to the occurrence of a given bigraph  
might not be unique. To ensure distinct decompositions, following
Krivine \etal~\cite{DBLP:journals/entcs/KrivineMT08}, we introduce the
following class of bigraphs.

\begin{definition}[solid]\label{def:solid}
  A bigraph is \emph{solid} if:
  \begin{itemize}
    \item All regions contain at least one node, and all outer names are connected to at least one link.
    \item No two sites or inner names are siblings
    \item No site has a region as a parent
    \item No outer name is linked to an inner name.
  \end{itemize}
\end{definition}

This definition is important when determining a suitable probability to apply a
rule.

\begin{definition}[reaction rule]\label{reaction_rule}
  A \emph{reaction rule} $\mathsf{R}$ is a pair of bigraphs
  $\mathsf{R} = (L, R)$, written $L \rrul R$, where $R$ and $L$ have
  the same interface and $L$ is solid.
\end{definition}

We also say $\mathsf{R}: L \rrul R$ is \emph{applicable} to $g$ iff $L$ occurs in $g$. 
In general, we are interested in applying a reaction rule within a larger
bigraph and as such provide the following reaction relation.

\begin{definition}[reaction relation]\label{reaction_relation}
  Given a reaction rule $\mathsf{R} : L \rrul R$, the \emph{reaction relation} 
  $\react_{{\mathsf{R}}}$ over ground bigraphs is defined by
  \begin{equation*}
  g \react_{{\mathsf{R}}} g' \qquad\text{iff}\qquad g = C \circ (L \otimes \id_{X}) \circ d \text{ and } g' = C \circ (R \otimes \id_{X}) \circ d
  \end{equation*}
  for some bigraph $C$, ground bigraph $d$, and set of names $X$.
\end{definition}

\begin{definition}[bigraphical reactive system (BRS)]\label{def:brs}
  A \emph{bigraphical reactive system} is a pair $(\mathcal{B}, \mathcal{R})$,
  where $\mathcal{B}$ is a set of ground bigraphs and  $\mathcal{R}$ is a set of reaction rules defined over  $\mathcal{B}$. It has reaction relation
  \begin{equation*}
  \react_{\mathcal{R}} = \bigcup\limits_{\mathsf{R} \in \mathcal{R}} \react_{\mathsf{R}}
  \end{equation*}
  which will be written $\react$ when $\mathcal{R}$ is understood.
\end{definition}

We indicate the set of reaction rules applicable to $g$ and yielding
$g'$ with $\mathcal{R}_{g \react g'}$. We also introduce the following
notation to count the \emph{concrete occurrences} of a reaction rule $\mathsf{R}$ from $g$ to $g'$
$$\rroccur{g}{g'} = |\{\conc{g''} \mid \conc{g} \react_{{\mathsf{R}}} \conc{g''} \text{ and } \conc{g''} \bumpeq \conc{g'}\}|$$
that is, we count how many concrete $\conc{g''}$ that are support equivalent to a concretion of $g'$ can be obtained by applying $\mathsf{R}$ to a concretion of $g$.

\begin{definition}[transition system]
A BRS with distinguished initial
bigraph $g_{0}  \in \mathcal{B}$ 
forms a \emph{transition system} with bigraphs as \emph{states} and state
 transitions defined by 
 generating all possible rewrites (reactions) in $\react$ from $g_{0}$ until we hit a fixed point (the set of all states $\mathcal{B}$).
\end{definition}

This transition system view is useful for defining probabilistic, stochastic,
and non-deterministic BRSs, where the transitions are assigned, for example,
specific probabilities.

Finally, we allow states to be labelled by \emph{bigraph predicates} which are also
specified as bigraphs~\cite{DBLP:journals/tochi/BenfordCRS16}. A state $g$
satisfies a predicate bigraph $P$ if $P$ occurs in $g$. These can be used to, for
example, identify invalid states for use in logical statements when performing
verification.

\subsection{Probabilistic Models}
\label{sec:probsystems}

In the following we assume basic familiarity with probability theory, see for
example~\cite{billingsley2012probability}.

Probabilistic systems can be described using Markov models/processes, where the
probability/rate of moving to a new state is based (strictly) on the current
state~\cite{kwiatkowska.ea_AdvancesAndChallengesOfProbabilisticMC:2010}.  A discrete time Markov chain (DTMC) labels each state transition
with a probability $0 \le p \le 1$ such that the sum of all outgoing edges from
a state is equal to $1$.
That is, a DTMC draws the \emph{next} state from a probability distribution of all possible states.

\begin{definition}[Probability Distribution]\label{def:defn_pdist} A probability distribution over a countable set $S$ is a function
  $\mu : S \to [0,1]$ satisfying $\sum_{s \in S} \mu(s) = 1$
\end{definition}

We use the notation $\mu = [ s_0 \mapsto p_0, s_{1} \mapsto p_{1}, \dots ]$ to
denote the distribution that chooses $s_0$ with probability $p_0$, and so on. We
assume all other states are chosen with probability 0.
To denote a set of probability distributions over $S$ we use $\mathcal{D}_{S}$,
dropping the subscript $S$ if it is clear from the context.
For verification purposes, we usually work with \emph{finite} probability
distributions with $S$ finite.

\begin{definition}[Discrete Time Markov Chain (DTMC)]\label{def:dtmc}
  A DTMC is a tuple $(S, s_{0}, P)$ where $S$ is a set of states, $s_{0} \in S$
  a distinguished initial state, and $P : S \to \mathcal{D}_{S}$ is a function
  assigning to each state $s \in S$ a probability distribution $\mu_{s}$ such
  that $\mu_{s}(s') : [0,1]$ is the transition probability from $s$ to $s'$.
\end{definition}

As distributions cannot be empty, each state $s \in S$ has at least one
transition. For terminal states $s_t$, we have $\mu_{s_{t}} = [ s_t \mapsto 1 ]$
-- the delta distribution.

To model continuous processes, we use continuous time Markov Chains (CTMCs)
that assign stochastic \emph{rates}, rather than probabilities, to state
transitions.

\begin{definition}[Continuous Time Markov Chain (CTMC)]\label{def:ctmc}
  A CTMC is a tuple $(S, s_{0}, R)$ where $S$ is a set of states, $s_{0} \in S$
  a distinguished initial state, and $R : S \times S \to \mathbb{R}_{\ge 0}$ the
  transition rate matrix matrix assigning a rate to each pair of states.
\end{definition}

A transition between $s$ and $s'$ can only occur if $R(s,s') > 0$, and if so the
probability of the transition occurring within time $t$ is modelled as an
exponential distribution, \ie $1 - e^{-R(s,s')t}$.
Unlike DTMCs, a CTMC allows terminal states where there is a 0 rate of
transitioning.

Markov decision processes~\cite{bellman1957markovian,howard1960dynamic} model
decision making in situations with both probabilistic outcomes and
non-deterministic decision making. Intuitively, an MDP extends a DTMC by
allowing a \emph{choice} of possible actions at each sate. Unlike a DTMC that
provides a single probability distribution per state, the choice of action
allows the multiple probability distributions per state.

\begin{definition}[Markov decision process (MDP)]\label{def:mdp}
  A MDP is a tuple \linebreak $(S, s_{0}, A, Step)$ where $S$ is a set of states, $s_{0}
  \in S$ a distinguished initial state, $A$ a set of \emph{actions}, and $Step :
  S \to 2^{A \times \mathcal{D}_{S}}$ 
  a function assigning to each state a set of
  possible actions with associated probability distributions.
\end{definition}

Unlike a DTMC, we allow states with no outgoing transitions, \ie $S \to \emptyset$.
When the choice of action for each step is fixed an MDP is a DTMC.

To allow practical analysis of probabilistic models it is useful to define
\emph{rewards} associated with being in a particular state.

\begin{definition}[state reward function]\label{def:state_rewards}
  For a DTMC, CTMC, or MDP, a \emph{state reward function} $r_{s} : S \to
  \mathbb{R}_{\ge 0}$ assigns to each state a reward. For states where rewards are not
  required $r_{s}$ maps the state to $0$.
\end{definition}

When working with bigraphs we associate rewards with bigraph predicates, allowing state
rewards to be defined as simply the sum of the rewards of all matching
predicates ($0$ if no predicates occur).

For MDPs we can also associate a reward for choosing a particular action.

\begin{definition}[action reward function]\label{def:action_rewards}
  An \emph{action reward structure} for an MDP $(S, s_{0}, A, Step)$ is a
  function $r_{a} : S \times A \to \mathbb{R}_{\ge 0}$ that assigns to each
  state, action pair a reward for performing that particular action. For actions
  where rewards are not required $r_{a}$ maps the action to $0$.
\end{definition}

Although we call these \emph{rewards}, they are often used to model
\emph{costs} associated with states/actions.

\section{Probabilistic Bigraphs}
\label{sec:pbrs}

Given a state (bigraph), we want to 
control the probability of moving into a given \emph{next} state (\ie a
bigraph). In other words,  we require a  
DTMC where the states resulting
from reactions are drawn from a probability distribution.

Our approach is similar to that of Bournez and
Hoyrup~\cite{bournez.hoyrup_RewritingLogicAndProbabilities:2003} who consider
abstract probabilistic rewrite systems. Here a \emph{weight} is assigned to each
rewrite rule which is then normalised based on which rules are applicable to a
given state. Other approaches to modelling probabilistic systems are possible,
for example, probabilistically determining which entities appear in the right-hand-side
of a rule. We discuss these further in \cref{sec:discussion}. Our approach allows re-use of
existing probabilistic model checking tools such a
PRISM~\cite{DBLP:conf/cav/KwiatkowskaNP11} or Storm~\cite{dehnert.ea_Storm:2017}
for analysis/verification.

\subsection{Probabilistic Bigraphical Reactive Systems}

A probabilistic BRS adds weights to standard reaction rules to determine transitions probabilities when defining the reaction relation.
This is, given a bigraph $g_{0}$ with $g_{0} \react g_{1}$ and $g_{0} \react g_{2}$ (for
arbitrary rules) we wish to choose $g_{1}$ and $g_{2}$ from a probability
distribution $\mu = [ g_{1} \mapsto p_{1}, g_{2} \mapsto p_{2} ]$, \ie
choose $g_{1}$ with probability $p_{1}$.

To account for multiple occurrences of a rule, we do not
directly specify probabilities for rules but instead assign a \emph{weight} which
is then normalised to a probability.

\begin{definition}[weighted reaction rule]\label{def:pbrs_rr}
  A \emph{weighted reaction rule} assigns to a reaction rule $L \rrul R$ a
  \emph{weight} $w$, $w \in \mathbb{R}_{\ge 0}$.
  We write weighted reaction rules as $L \rrulp{w} R$.
\end{definition}

The \emph{weight} determines how likely a particular rule is to be applied
\emph{relative} to all other (applicable) rules. Rules with weight $w=0$ are never applied. In the following, we write
$\react_{\mathcal{R}}$ to indicate that a set of weighted reaction rules
$\mathcal{R}$ is treated as a set of standard reaction rules (see
Definition~\ref{def:brs}) by dropping all the weights.

\begin{definition}[total weight]
Given a set of weighted reaction rules $\mathcal{R}$, the \emph{total weight from $g$ to $g'$} is
$$\totw{g}{g'} = \sum_{\mathsf{R}\in \mathcal{R}_{g \react g'}} w_{\mathsf{R}} \cdot \rroccur{g}{g'}$$ where $w_{\mathsf{R}}$ is the
weight of reaction rule $\mathsf{R}$. Given a set of ground bigraphs
$\mathcal{B}$, the \emph{total weight from $g$} is
$$\totw{g}{\_} = \sum_{g'\in \mathcal{B}} \totw{g}{g'}$$\end{definition}

\begin{definition}[reaction probability distribution]\label{pbrs_rpd}
Given a set of ground bigraphs $\mathcal{B}$, a set of weighted reaction
rules $\mathcal{R}$ and $g \in \mathcal{B}$, the \emph{reaction probability distribution from $g$} is
$$\mu_g = [g_0 \mapsto p_0 , g_1 \mapsto p_1, \ldots]\qquad \text{with} \qquad p_i = \frac{\totw{g}{g_i}}{\totw{g}{\_}}$$
for every $g_i\in \mathcal{B}$ such that $g \react_{\mathcal{R}} g_i$. If there are no such $g_i$s, then $\mu_g = [ g \mapsto 1 ]$.
\end{definition}

Reaction probability distributions are then used to define a probabilistic reaction
relation over ground bigraphs.

\begin{definition}[probabilistic bigraphical reactive system (PBRS)]\label{pbrs_defn}
  A \emph{probabilistic BRS} is a pair $(\mathcal{B}, \mathcal{R})$, where
  $\mathcal{B}$ is a set of ground bigraphs, and $\mathcal{R}$ is a set of
  weighted reaction rules. It has \emph{probabilistic reaction relation} defined by
  $$g \reactp{p} g' \qquad\text{iff}\qquad \mu_g(g')=p$$
  with $g,g'\in \mathcal{B}$.
\end{definition}

The correspondence between PBRS and DTMC is as follows:

\begin{lemma}
  A PBRS $(\mathcal{B}, \mathcal{R})$ is a DTMC
  $(S, s_{0}, P : S \to \mathcal{D}_{S})$.
\begin{proof}
  Take $S = \mathcal{B}$,
  $s_{0} =  g_{0} \in \mathcal{B}$, and $P(s) = \mu_{g}$ (\cref{pbrs_rpd}) for $s \in S$, $g \in \mathcal{B}$, and $s=g$.\end{proof}

\end{lemma}

\mscom{Also double check the implementation in BigrphER}

From a practical standpoint, the use of weighted reaction rules  allows
modelling only the \emph{relative} probability a particular rule is executed.
Unfortunately, this makes it difficult to specify an \emph{exact} probability
between states. Doing so is often impractical, requiring significant effort to
control the applicable rules and number of occurrences such that the normalised
probabilities are exact. Usually the relative outcomes are what is important,
and so far we have not encountered any situation where this is a particular issue.

\subsection{Example PBRS}
\label{sec:example_pbrs}
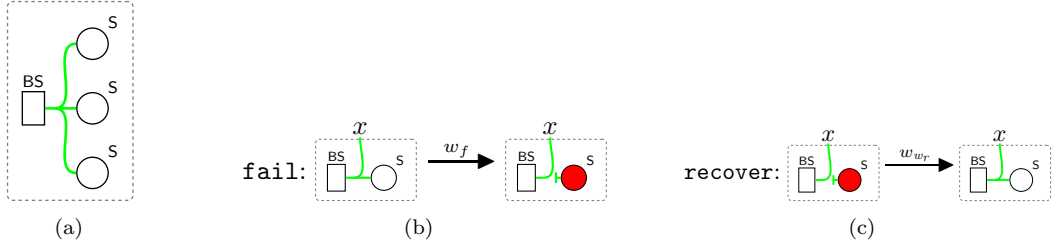
\begin{figure}
  \begin{subfigure}[t]{0.2\linewidth}
  \centering
  \resizebox{0.6\linewidth}{!}{
    \begin{tikzpicture}
      \node[draw,minimum height=10] (bs) {};
      \node[lbl, anchor=south] at (bs.north) (bs_lbl) {BS};

      \node[draw,circle] at ($(bs.north east) + (0.5,0.5)$) (n1) {};
      \node[lbl, anchor=south west] at (n1.north east) (n1_lbl) {S};
      \node[draw,circle] at ($(bs.east) + (0.5,0)$) (n2) {};
      \node[lbl, anchor=south west] at (n2.north east) (n2_lbl) {S};
      \node[draw,circle] at ($(bs.south east) + (0.5,-0.5)$) (n3) {};
      \node[lbl, anchor=south west] at (n3.north east) (n3_lbl) {S};

      \node[big region, fit=(bs)(bs_lbl)(n1)(n1_lbl)(n2)(n2_lbl)(n3)(n3_lbl)]
      (r1) {};

\coordinate (h1) at ($(bs.east) + (0.1,0)$);
      \draw[big edge] (bs.east) to[] (h1);
      \draw[big edge] (n1.west) to[out=180, in=0] (h1);
      \draw[big edge] (n2.west) to[out=180, in=0] (h1);
      \draw[big edge] (n3.west) to[out=180, in=0] (h1);
    \end{tikzpicture}
  }
  \caption{}
  \label{fig:bg-example-wsn-b}
\end{subfigure}
\begin{subfigure}[t]{0.35\linewidth}
  \centering
  \resizebox{0.9\linewidth}{!}{
    \begin{tikzpicture}
      \begin{scope}[local bounding box=lhs]
        \node[draw,minimum height=10] (bs) {};
        \node[lbl, anchor=south] at (bs.north) (bs_lbl) {BS};

        \node[draw,circle] at ($(bs.east) + (0.5,0)$) (n1) {};
        \node[lbl, anchor=south west] at (n1.north east) (n1_lbl) {S};

        \node[big region, fit=(bs)(bs_lbl)(n1)(n1_lbl)]
        (r1) {};

        \node[inner sep=1] at ($($(bs)!0.5!(n1)$) + (0,0.65)$) (x) {$x$};

\coordinate (h1) at ($(bs.east) + (0.1,0)$);
        \draw[big edge] (bs.east) to[] (h1);
        \draw[big edge] (n1.west) to[out=180, in=0] (h1);

        \draw[big edge] (h1) to[out=0, in=-90] (x);
      \end{scope}

      \begin{scope}[shift={(2.5,0)}, local bounding box=rhs]
        \node[draw,minimum height=10] (bs) {};
        \node[lbl, anchor=south] at (bs.north) (bs_lbl) {BS};

        \node[draw,circle, fill=red] at ($(bs.east) + (0.5,0)$) (n1) {};
        \node[lbl, anchor=south west] at (n1.north east) (n1_lbl) {S};

        \node[big region, fit=(bs)(bs_lbl)(n1)(n1_lbl)]
        (r1) {};

        \node[inner sep=1] at ($($(bs)!0.5!(n1)$) + (0,0.65)$) (x) {$x$};

\coordinate (h1) at ($(bs.east) + (0.1,0)$);
        \draw[big edge] (bs.east) to[] (h1);
        \draw[big edgec] (n1.west) to[out=180, in=0] ($(h1) + (0.15, 0)$);

        \draw[big edge] (h1) to[out=0, in=-90] (x);
      \end{scope}

      \node[] at ($(lhs.east)!0.5!(rhs.west)$) {$\rrulp{\wfail}$};
      \node[anchor=east, yshift=-4] at (lhs.west) {\texttt{fail}:};
    \end{tikzpicture}
  }
  \caption{}
  \label{fig:bg-example-wsn-failure}
\end{subfigure}
\begin{subfigure}[t]{0.35\linewidth}
  \centering
  \resizebox{0.9\linewidth}{!}{
    \begin{tikzpicture}
      \begin{scope}[local bounding box=lhs]
        \node[draw,minimum height=10] (bs) {};
        \node[lbl, anchor=south] at (bs.north) (bs_lbl) {BS};

        \node[draw,circle, fill=red] at ($(bs.east) + (0.5,0)$) (n1) {};
        \node[lbl, anchor=south west] at (n1.north east) (n1_lbl) {S};

        \node[big region, fit=(bs)(bs_lbl)(n1)(n1_lbl)]
        (r1) {};

        \node[inner sep=1] at ($($(bs)!0.5!(n1)$) + (0,0.65)$) (x) {$x$};

\coordinate (h1) at ($(bs.east) + (0.1,0)$);
        \draw[big edge] (bs.east) to[] (h1);
\draw[big edgec] (n1.west) to[out=180, in=0] ($(h1) + (0.15, 0)$);

        \draw[big edge] (h1) to[out=0, in=-90] (x);
      \end{scope}

      \begin{scope}[shift={(2.5,0)}, local bounding box=rhs]
        \node[draw,minimum height=10] (bs) {};
        \node[lbl, anchor=south] at (bs.north) (bs_lbl) {BS};

        \node[draw,circle] at ($(bs.east) + (0.5,0)$) (n1) {};
        \node[lbl, anchor=south west] at (n1.north east) (n1_lbl) {S};

        \node[big region, fit=(bs)(bs_lbl)(n1)(n1_lbl)]
        (r1) {};

        \node[inner sep=1] at ($($(bs)!0.5!(n1)$) + (0,0.65)$) (x) {$x$};

\coordinate (h1) at ($(bs.east) + (0.1,0)$);
        \draw[big edge] (bs.east) to[] (h1);
        \draw[big edge] (n1.west) to[out=180, in=0] (h1);

        \draw[big edge] (h1) to[out=0, in=-90] (x);
      \end{scope}

      \node[] at ($(lhs.east)!0.5!(rhs.west)$) {$\rrulp{w_\mathit\wcon}$};
      \node[anchor=east, yshift=-4] at (lhs.west) {\texttt{recover}: };
    \end{tikzpicture}
  }
  \caption{}
  \label{fig:bg-example-wsn-recover}
\end{subfigure}
\caption{Wireless sensor networks as bigraphs. (a) Bigraph representing a
  WSN with base-station \textsf{BS} and three sensors \textsf{S}.
  (b) Probabilistic reaction rule modelling failure of a sensor. (c)
  Probabilistic reaction rule modelling recovery of a sensor.}
\label{fig:bg-example-wsn}
\end{figure}

 Consider a Wireless Sensor Network (WSN) with three sensor nodes
(\textsf{S}) and a base-station (\textsf{BS}), as shown in 
  \cref{fig:bg-example-wsn}. The base station is represented by the rectangle and the three sensors are represented by circles.  There is a link between the base station and the sensors. Due to hostile deployment environments, sensors
often fail. We model failure using the reaction rule \texttt{fail}
  that marks a sensor as failed (red circle) and unlinks
it from the base-station. The  rule \texttt{recover}
 allows a failed sensor (red circle) to re-connect with
the base-station.
  \texttt{fail} (b) has weight $\wfail$ and \texttt{recover} (c)
has weight $\wcon$. Note that while \texttt{fail} and \texttt{recover} are
behaviourally \emph{inverse} rules, their weights differ.

\begin{figure}
  \centering
  \resizebox{0.6\linewidth}{!}{
    \begin{tikzpicture}
      \begin{scope}[local bounding box=init]
        \node[draw,minimum height=10] (bs) {};
        \node[lbl, anchor=south] at (bs.north) (bs_lbl) {BS};

        \node[draw,circle] at ($(bs.north east) + (0.5,0.5)$) (n1) {};
        \node[lbl, anchor=south west] at (n1.north east) (n1_lbl) {S};
        \node[draw,circle] at ($(bs.east) + (0.5,0)$) (n2) {};
        \node[lbl, anchor=south west] at (n2.north east) (n2_lbl) {S};
        \node[draw,circle] at ($(bs.south east) + (0.5,-0.5)$) (n3) {};
        \node[lbl, anchor=south west] at (n3.north east) (n3_lbl) {S};

        \node[big region, fit=(bs)(bs_lbl)(n1)(n1_lbl)(n2)(n2_lbl)(n3)(n3_lbl)]
        (r1) {};

\coordinate (h1) at ($(bs.east) + (0.1,0)$);
        \draw[big edge] (bs.east) to[] (h1);
        \draw[big edge] (n1.west) to[out=180, in=0] (h1);
        \draw[big edge] (n2.west) to[out=180, in=0] (h1);
        \draw[big edge] (n3.west) to[out=180, in=0] (h1);
      \end{scope}
      \begin{scope}[shift={(2.5,0)}, local bounding box=s1]
        \node[draw,minimum height=10] (bs) {};
        \node[lbl, anchor=south] at (bs.north) (bs_lbl) {BS};

        \node[draw,circle, fill=red] at ($(bs.north east) + (0.5,0.5)$) (n1) {};
        \node[lbl, anchor=south west] at (n1.north east) (n1_lbl) {S};
        \node[draw,circle] at ($(bs.east) + (0.5,0)$) (n2) {};
        \node[lbl, anchor=south west] at (n2.north east) (n2_lbl) {S};
        \node[draw,circle] at ($(bs.south east) + (0.5,-0.5)$) (n3) {};
        \node[lbl, anchor=south west] at (n3.north east) (n3_lbl) {S};

        \node[big region, fit=(bs)(bs_lbl)(n1)(n1_lbl)(n2)(n2_lbl)(n3)(n3_lbl)]
        (r1) {};

\coordinate (h1) at ($(bs.east) + (0.1,0)$);
        \draw[big edge] (bs.east) to[] (h1);
        \draw[big edgec] (n1.west) to[out=180, in=0] ($(n1.west) + (-0.15, 0)$);
        \draw[big edge] (n2.west) to[out=180, in=0] (h1);
        \draw[big edge] (n3.west) to[out=180, in=0] (h1);
      \end{scope}
      \begin{scope}[shift={(5,0)}, local bounding box=s2]
        \node[draw,minimum height=10] (bs) {};
        \node[lbl, anchor=south] at (bs.north) (bs_lbl) {BS};

        \node[draw,circle,fill=red] at ($(bs.north east) + (0.5,0.5)$) (n1) {};
        \node[lbl, anchor=south west] at (n1.north east) (n1_lbl) {S};
        \node[draw,circle, fill=red] at ($(bs.east) + (0.5,0)$) (n2) {};
        \node[lbl, anchor=south west] at (n2.north east) (n2_lbl) {S};
        \node[draw,circle] at ($(bs.south east) + (0.5,-0.5)$) (n3) {};
        \node[lbl, anchor=south west] at (n3.north east) (n3_lbl) {S};

        \node[big region, fit=(bs)(bs_lbl)(n1)(n1_lbl)(n2)(n2_lbl)(n3)(n3_lbl)]
        (r1) {};

\coordinate (h1) at ($(bs.east) + (0.1,0)$);
        \draw[big edge] (bs.east) to[] (h1);
        \draw[big edgec] (n1.west) to[out=180, in=0] ($(n1.west) + (-0.15, 0)$);
        \draw[big edgec] (n2.west) to[out=180, in=0] ($(n2.west) + (-0.10, 0)$);
        \draw[big edge] (n3.west) to[out=180, in=0] (h1);
      \end{scope}
      \begin{scope}[shift={(7.5,0)}, local bounding box=s3]
        \node[draw,minimum height=10] (bs) {};
        \node[lbl, anchor=south] at (bs.north) (bs_lbl) {BS};

        \node[draw,circle,fill=red] at ($(bs.north east) + (0.5,0.5)$) (n1) {};
        \node[lbl, anchor=south west] at (n1.north east) (n1_lbl) {S};
        \node[draw,circle,fill=red] at ($(bs.east) + (0.5,0)$) (n2) {};
        \node[lbl, anchor=south west] at (n2.north east) (n2_lbl) {S};
        \node[draw,circle,fill=red] at ($(bs.south east) + (0.5,-0.5)$) (n3) {};
        \node[lbl, anchor=south west] at (n3.north east) (n3_lbl) {S};

        \node[big region, fit=(bs)(bs_lbl)(n1)(n1_lbl)(n2)(n2_lbl)(n3)(n3_lbl)]
        (r1) {};

\coordinate (h1) at ($(bs.east) + (0.1,0)$);
        \draw[big edgec] (bs.east) to[] (h1);
        \draw[big edgec] (n1.west) to[out=180, in=0] ($(n1.west) + (-0.15, 0)$);
        \draw[big edgec] (n2.west) to[out=180, in=0] ($(n2.west) + (-0.10, 0)$);
        \draw[big edgec] (n3.west) to[out=180, in=0] ($(n3.west) + (-0.15, 0)$);
      \end{scope}

\draw[-latex] (init.50) to[out=45,in=150] node[above] {\tiny $1$} (s1.130);
      \draw[-latex] (s1.50) to[out=45,in=150] node[above] {\tiny $\frac{2\wfail}{2\wfail + \wcon}$} (s2.130);
      \draw[-latex] (s2.50) to[out=45,in=150] node[above] {\tiny $\frac{\wfail}{\wfail + 2\wcon}$} (s3.130);

      \draw[-latex] (s3.220) to[out=-120,in=-50] node[below] {\tiny $1$} (s2.-40);
      \draw[-latex] (s2.220) to[out=-120,in=-50] node[below] {\tiny $\frac{2\wcon}{\wfail + 2\wcon}$} (s1.-40);
      \draw[-latex] (s1.220) to[out=-120,in=-50] node[below] {\tiny $\frac{\wcon}{2\wfail + \wcon}$} (init.-40);

    \end{tikzpicture}
  }
      \caption{DTMC for bigraph model of \cref{fig:bg-example-wsn}. States are bigraphs $\mathit{g_0} \dots \mathit{g_3}$ from left to right.}
  \label{fig:bg-example-wsn-DTMC}
\end{figure}
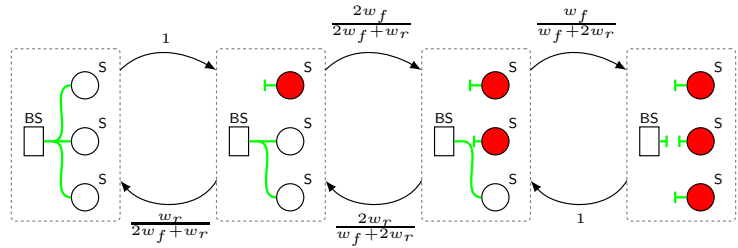

\begin{figure}
  \centering
  \resizebox{0.25\linewidth}{!}{
    \begin{tikzpicture}
      \begin{scope}[local bounding box=init]
        \node[draw,minimum height=10] (bs) {};
        \node[lbl conc, anchor=south] at (bs.north) (bs_lbl) {$v_0$};

        \node[draw,circle] at ($(bs.north east) + (0.5,0.5)$) (n1) {};
        \node[lbl conc, anchor=south west] at (n1.north east) (n1_lbl) {$v_1$};
        \node[draw,circle] at ($(bs.east) + (0.5,0)$) (n2) {};
        \node[lbl conc, anchor=south west] at (n2.north east) (n2_lbl) {$v_2$};
        \node[draw,circle] at ($(bs.south east) + (0.5,-0.5)$) (n3) {};
        \node[lbl conc, anchor=south west] at (n3.north east) (n3_lbl) {$v_3$};

        \node[big region, fit=(bs)(bs_lbl)(n1)(n1_lbl)(n2)(n2_lbl)(n3)(n3_lbl)]
        (r1) {};

\coordinate (h1) at ($(bs.east) + (0.1,0)$);
        \draw[big edge] (bs.east) to[] (h1);
        \draw[big edge] (n1.west) to[out=180, in=0] (h1);
        \draw[big edge] (n2.west) to[out=180, in=0] (h1);
        \draw[big edge] (n3.west) to[out=180, in=0] (h1);
      \end{scope}

      \begin{scope}[shift={(2.5,2.5)}, local bounding box=s1]
        \node[draw,minimum height=10] (bs) {};
        \node[lbl conc, anchor=south] at (bs.north) (bs_lbl) {$v_0$};

        \node[draw,circle, fill=red] at ($(bs.north east) + (0.5,0.5)$) (n1) {};
        \node[lbl conc, anchor=south west] at (n1.north east) (n1_lbl) {$v_1$};
        \node[draw,circle] at ($(bs.east) + (0.5,0)$) (n2) {};
        \node[lbl conc, anchor=south west] at (n2.north east) (n2_lbl) {$v_2$};
        \node[draw,circle] at ($(bs.south east) + (0.5,-0.5)$) (n3) {};
        \node[lbl conc, anchor=south west] at (n3.north east) (n3_lbl) {$v_3$};

        \node[big region, fit=(bs)(bs_lbl)(n1)(n1_lbl)(n2)(n2_lbl)(n3)(n3_lbl)]
        (r1) {};

\coordinate (h1) at ($(bs.east) + (0.1,0)$);
        \draw[big edge] (bs.east) to[] (h1);
\draw[big edgec] (n1.west) to[out=180, in=0] ($(n1.west) + (-0.15, 0)$);
        \draw[big edge] (n2.west) to[out=180, in=0] (h1);
        \draw[big edge] (n3.west) to[out=180, in=0] (h1);
\end{scope}

      \begin{scope}[shift={(2.5,0)}, local bounding box=s2]
        \node[draw,minimum height=10] (bs) {};
        \node[lbl conc, anchor=south] at (bs.north) (bs_lbl) {$v_0$};

        \node[draw,circle] at ($(bs.north east) + (0.5,0.5)$) (n1) {};
        \node[lbl conc, anchor=south west] at (n1.north east) (n1_lbl) {$v_1$};
        \node[draw,circle, fill=red] at ($(bs.east) + (0.5,0)$) (n2) {};
        \node[lbl conc, anchor=south west] at (n2.north east) (n2_lbl) {$v_2$};
        \node[draw,circle] at ($(bs.south east) + (0.5,-0.5)$) (n3) {};
        \node[lbl conc, anchor=south west] at (n3.north east) (n3_lbl) {$v_3$};

        \node[big region, fit=(bs)(bs_lbl)(n1)(n1_lbl)(n2)(n2_lbl)(n3)(n3_lbl)]
        (r1) {};

\coordinate (h1) at ($(bs.east) + (0.1,0)$);
        \draw[big edge] (bs.east) to[] (h1);
        \draw[big edge] (n1.west) to[out=180, in=0] (h1);
        \draw[big edgec] (n2.west) to[out=180, in=0] ($(n2.west) + (-0.10, 0)$);
        \draw[big edge] (n3.west) to[out=180, in=0] (h1);
      \end{scope}
      \begin{scope}[shift={(2.5,-2.5)}, local bounding box=s3]
        \node[draw,minimum height=10] (bs) {};
        \node[lbl conc, anchor=south] at (bs.north) (bs_lbl) {$v_0$};

        \node[draw,circle] at ($(bs.north east) + (0.5,0.5)$) (n1) {};
        \node[lbl conc, anchor=south west] at (n1.north east) (n1_lbl) {$v_1$};
        \node[draw,circle] at ($(bs.east) + (0.5,0)$) (n2) {};
        \node[lbl conc, anchor=south west] at (n2.north east) (n2_lbl) {$v_2$};
        \node[draw,circle, fill=red] at ($(bs.south east) + (0.5,-0.5)$) (n3) {};
        \node[lbl conc, anchor=south west] at (n3.north east) (n3_lbl) {$v_3$};

        \node[big region, fit=(bs)(bs_lbl)(n1)(n1_lbl)(n2)(n2_lbl)(n3)(n3_lbl)]
        (r1) {};

\coordinate (h1) at ($(bs.east) + (0.1,0)$);
        \draw[big edge] (bs.east) to[] (h1);

        \draw[big edge] (n1.west) to[out=180, in=0] (h1);
        \draw[big edgec] (n3.west) to[out=180, in=0] ($(n3.west) + (-0.15, 0)$);
        \draw[big edge] (n2.west) to[out=180, in=0] (h1);
\end{scope}

\draw[-latex] (init.50) to[out=0,in=180] node[right] {} (s1.west);
      \draw[-latex] (init.east) to[out=0,in=180] node[above] {} (s2.west);
      \draw[-latex] (init.-50) to[out=0,in=180] node[right] {} (s3.west);
    \end{tikzpicture}
  }
  \caption{Concrete occurrences of reaction rule $\mathtt{fail}$ from $g_0$ to $g_1$ (see \cref{fig:bg-example-wsn}).}
  \label{fig:bg-example-wsn-support}
\end{figure}
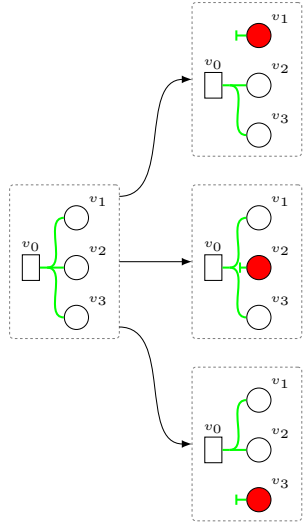

The resulting transition system for this WSN is in \cref{fig:bg-example-wsn-DTMC}. From the initial state $\mathit{g_0}$, we   determine the distribution of next states
$\mu_{\mathit{g_0}}$. 
In this case, reaction rule $\mathtt{recover}$ is not applicable and we can only apply $\mathtt{fail}$.
It may be surprising that even though there are three sensors, the probability of transitioning to $g_1$ is $\mu_{g_0}(g_1)=1$.
This is because support equivalent (concrete) bigraphs are combined when computing the states resulting from the application of a reaction rule. 
To see this more clearly, we show explicitly the three concrete occurrences of $\mathtt{fail}$ from $g_0$ to $g_1$ in
\cref{fig:bg-example-wsn-support}.
The key observation is that through renamings
$v_{2} \to v_{1}$ and $v_{3} \to v_{1}$, these concrete bigraphs are support equivalent and therefore they correspond to the single abstract state $g_1$. 
Hence, we have 
$\rroccurr{g_0}{g_1}{\mathtt{fail}} = 3$, $\totw{g_0}{g_1} = 3 \wfail$, and $\totw{g_0}{\_} = 3 \wfail$, giving the overall reaction probability of 1.

In state $g_1$, with one failed sensor, we have
$\rroccurr{g_1}{g_2}{\mathtt{fail}} = 2$ and
$\rroccurr{g_1}{g_0}{\mathtt{con}} = 1$. 
Normalising this over the total weight $\totw{g_1}{\_} = 2 \wfail + \wcon$
we obtain a
$\frac{2\wfail}{2\wfail + \wcon}$ probability
that another sensor fails, and a probability
$\frac{\wcon}{2\wfail + \wcon}$ that the failed
sensor recovers. 

Importantly, due to support equivalence, transition probability corresponds to the probability that \textbf{any} sensor fails
rather than the probability that a \textbf{particular} sensor fails
(\ie $\frac{\wfail}{2\wfail + \wcon}$).

This process of normalising weights to probabilities continues until we obtain the full DTMC as shown.

\section{Action Bigraphs}
\label{sec:mdps}

PBRSs allow a single distribution of possible next states defined over
$\emph{all}$ rules $\mathcal{R}$. However, for systems such as controllers, we
want actions taken by the controller to affect the possible evolution of the
system, \eg by restricting the reaction rules. That is, we want multiple
distributions that are determined by the action taken. To this end, we introduce
\emph{action bigraphical reactive systems} (ABRS), where the resulting
transition system is a Markov decision processes (MDP -- \cref{def:mdp}).

ABRSs extend PBRSs by
allowing a \emph{choice} of probability distributions at each step. We call such choices \emph{actions}.

\begin{definition}[Action]\label{def:nbrs_action}
  An \emph{action} is a non-empty set of weighted reaction rules (\cref{def:pbrs_rr})
  that determines the rewrites that can be performed \emph{if} the
  action is chosen. We say an action is \emph{applicable} to a bigraph $g$ if at
  least one rule from the action is applicable to $g$.
\end{definition}

As actions are simply sets of rules, the same reaction rule may appear in multiple actions if required, \eg if two different control actions allow updating of the same state.
We use the notation $\restriction_{A}$ to restrict definitions to consider only rules in $A$. For example $\totwr{g}{g_i}{A}$ is the total weight between states $g$ and $g_i$ when considering only rules in $A \subseteq \mathcal{R}$ rather than in $\mathcal{R}$.

To move from weighted to probabilistic rules, we apply, individually for each action,
the normalisation procedure from PBRSs, \ie

$$p_i = \frac{\totwr{g}{g_i}{A_{}}}{\totwr{g}{\_}{A_{}}}$$

After normalising, we obtain a
set of probability distributions; one for each applicable action, allowing us to construct the MDP transition function $Step : S \to 2^{A \times \mathcal{D}_{S}}$.

We then define an Action BRS as follows.

\begin{definition}[Action BRS (ABRS)]\label{nbrs_defn}
  An \emph{Action BRS} is a triple $(\mathcal{B}, \mathcal{R},
  \mathcal{A})$, where $\mathcal{B}$ is a set of (ground) bigraphs, $\mathcal{R}$ is a set of weighted reaction rules
  over $\mathcal{B}$, and $\mathcal{A} = \{ A_{0} \subseteq \mathcal{R}, \dots,
  A_{n} \subseteq \mathcal{R} \}$ is a set of actions.
  It has a \emph{reaction relation} defined by
  $$g \reactp{(A_i, p)} g' \qquad\text{iff}\qquad \distr{g}{g'}{A_{i}}=p$$ for
  each applicable action $A_i \in \mathcal{A}$ with $g,g'\in \mathcal{B}$.
\end{definition}

Just as a PBRS is a DTMC, an ABRS is an MDP:

\begin{lemma}
  An ABRS $(\mathcal{B}, \mathcal{R}, \mathcal{A})$ is an MDP
  $(S, s_{0}, A, Step : S \to 2^{A \times \mathcal{D}_{S}})$.
  \begin{proof}
    Take $S = \mathcal{B}$, $s_{0} = g_{0} \in \mathcal{B}$, $A = \mathcal{A}$, and for $s=g$ with $s\in S$ and $g \in \mathcal{B}$ define
    $$
    Step(s) =
    \begin{cases}
    \{ (A_{i}, \distrf{g}{A_{i}}), \dots, ( A_{n}, \distrf{g}{A_{n}}) \} & \mbox{for all $A_i \in \mathcal{A}$ applicable to $g$}\\
    \emptyset & \mbox{if no action applies to $g$}
    \end{cases}
    $$
\end{proof}
\end{lemma}

As with MDPs, we can assign rewards (\cref{def:action_rewards}) for choosing a
particular action to allow optimisation of decision processes.

Like MDPs, ABRS allow terminal states (the empty set of distributions) if there is no
applicable action, however, for practical analysis, \eg in PRISM, we usually
require at least one action per state\mscom{Do we also enforce this in BigraphER?}. In a similar manner to PBRS, in the case
no \emph{action} applies, we can add an trivial action containing the
identity reaction for the current state.

\subsection{Example ABRS}

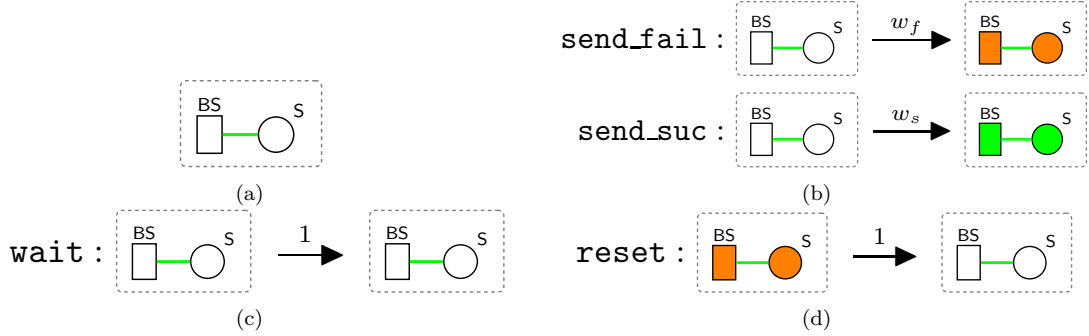
\begin{figure}
  \centering
  \begin{subfigure}[t]{0.45\linewidth}
    \centering
    \resizebox{0.3\linewidth}{!}{
      \begin{tikzpicture}
        \node[draw, minimum height=10] (bs) {};
        \node[lbl, anchor=south] at (bs.north) (bs_lbl) {BS};
        \node[draw,circle] at ($(bs.east) + (0.5,0)$) (n1) {};
        \node[lbl, anchor=south west] at (n1.north east) (n1_lbl) {S};
        \draw[big edge] (bs.east) to[out=0, in=180] (n1);
        \node[big region, fit=(bs)(bs_lbl)(n1)(n1_lbl)] (r1) {};
      \end{tikzpicture}
    }
    \caption{}
    \label{fig:nbrs-example-init}
  \end{subfigure}
  \begin{subfigure}[t]{0.45\linewidth}
    \centering
    \resizebox{\linewidth}{!}{
      \begin{tikzpicture}
        \begin{scope}
          \begin{scope}[local bounding box=lhs]
            \node[draw, minimum height=10] (bs) {};
            \node[lbl, anchor=south] at (bs.north) (bs_lbl) {BS};
            \node[draw,circle] at ($(bs.east) + (0.5,0)$) (n1) {};
            \node[lbl, anchor=south west] at (n1.north east) (n1_lbl) {S};
            \draw[big edge] (bs.east) to[out=0, in=180] (n1);
            \node[big region, fit=(bs)(bs_lbl)(n1)(n1_lbl)] (r1) {};
          \end{scope}
          \begin{scope}[shift={(2.5,0)}, local bounding box=rhs]
            \node[draw, minimum height=10, fill=orange] (bs) {};
            \node[lbl, anchor=south] at (bs.north) (bs_lbl) {BS};
            \node[draw,circle, fill=orange] at ($(bs.east) + (0.5,0)$) (n1) {};
            \node[lbl, anchor=south west] at (n1.north east) (n1_lbl) {S};
            \draw[big edge] (bs.east) to[out=0, in=180] (n1);
            \node[big region, fit=(bs)(bs_lbl)(n1)(n1_lbl)] (r1) {};
          \end{scope}
          \node[] at ($(lhs.east)!0.5!(rhs.west)$) {$\rrulp{\wfail}$};
          \node[anchor=east] at (lhs.west) (lbl) {$\mathtt{send\_fail:}$};
        \end{scope}
        \begin{scope}[shift={(0,-1)}]
          \begin{scope}[local bounding box=lhs]
            \node[draw, minimum height=10] (bs) {};
            \node[lbl, anchor=south] at (bs.north) (bs_lbl) {BS};
            \node[draw,circle] at ($(bs.east) + (0.5,0)$) (n1) {};
            \node[lbl, anchor=south west] at (n1.north east) (n1_lbl) {S};
            \draw[big edge] (bs.east) to[out=0, in=180] (n1);
            \node[big region, fit=(bs)(bs_lbl)(n1)(n1_lbl)] (r1) {};
          \end{scope}
          \begin{scope}[shift={(2.5,0)}, local bounding box=rhs]
            \node[draw, minimum height=10, fill=green] (bs) {};
            \node[lbl, anchor=south] at (bs.north) (bs_lbl) {BS};
            \node[draw,circle, fill=green] at ($(bs.east) + (0.5,0)$) (n1) {};
            \node[lbl, anchor=south west] at (n1.north east) (n1_lbl) {S};
            \draw[big edge] (bs.east) to[out=0, in=180] (n1);
            \node[big region, fit=(bs)(bs_lbl)(n1)(n1_lbl)] (r1) {};
          \end{scope}
          \node[] at ($(lhs.east)!0.5!(rhs.west)$) {$\rrulp{\wsuc}$};
          \node[anchor=east] at (lhs.west) (lbl) {$\mathtt{send\_suc:}$};
        \end{scope}
      \end{tikzpicture}
    }
    \caption{}
    \label{fig:nbrs-example-send}
  \end{subfigure}

  \begin{subfigure}[t]{0.45\linewidth}
    \centering
    \resizebox{0.95\linewidth}{!}{
      \begin{tikzpicture}
        \begin{scope}
          \begin{scope}[local bounding box=lhs]
            \node[draw, minimum height=10] (bs) {};
            \node[lbl, anchor=south] at (bs.north) (bs_lbl) {BS};
            \node[draw,circle] at ($(bs.east) + (0.5,0)$) (n1) {};
            \node[lbl, anchor=south west] at (n1.north east) (n1_lbl) {S};
            \draw[big edge] (bs.east) to[out=0, in=180] (n1);
            \node[big region, fit=(bs)(bs_lbl)(n1)(n1_lbl)] (r1) {};
          \end{scope}
          \begin{scope}[shift={(2.5,0)}, local bounding box=rhs]
            \node[draw, minimum height=10] (bs) {};
            \node[lbl, anchor=south] at (bs.north) (bs_lbl) {BS};
            \node[draw,circle] at ($(bs.east) + (0.5,0)$) (n1) {};
            \node[lbl, anchor=south west] at (n1.north east) (n1_lbl) {S};
            \draw[big edge] (bs.east) to[out=0, in=180] (n1);
            \node[big region, fit=(bs)(bs_lbl)(n1)(n1_lbl)] (r1) {};
          \end{scope}
          \node[] at ($(lhs.east)!0.5!(rhs.west)$) {$\rrulp{1}$};
          \node[anchor=east] at (lhs.west) (lbl) {$\mathtt{wait:}$};
        \end{scope}
      \end{tikzpicture}
    }
    \caption{}
    \label{fig:nbrs-example-wait}
  \end{subfigure}
  \begin{subfigure}[t]{0.45\linewidth}
    \centering
    \resizebox{0.95\linewidth}{!}{
      \begin{tikzpicture}
        \begin{scope}
          \begin{scope}[local bounding box=lhs]
            \node[draw, minimum height=10, fill=orange] (bs) {};
            \node[lbl, anchor=south] at (bs.north) (bs_lbl) {BS};
            \node[draw,circle, fill=orange] at ($(bs.east) + (0.5,0)$) (n1) {};
            \node[lbl, anchor=south west] at (n1.north east) (n1_lbl) {S};
            \draw[big edge] (bs.east) to[out=0, in=180] (n1);
            \node[big region, fit=(bs)(bs_lbl)(n1)(n1_lbl)] (r1) {};
          \end{scope}
          \begin{scope}[shift={(2.5,0)}, local bounding box=rhs]
            \node[draw, minimum height=10] (bs) {};
            \node[lbl, anchor=south] at (bs.north) (bs_lbl) {BS};
            \node[draw,circle] at ($(bs.east) + (0.5,0)$) (n1) {};
            \node[lbl, anchor=south west] at (n1.north east) (n1_lbl) {S};
            \draw[big edge] (bs.east) to[out=0, in=180] (n1);
            \node[big region, fit=(bs)(bs_lbl)(n1)(n1_lbl)] (r1) {};
          \end{scope}
          \node[] at ($(lhs.east)!0.5!(rhs.west)$) {$\rrulp{1}$};
          \node[anchor=east] at (lhs.west) (lbl) {$\mathtt{reset:}$};
        \end{scope}
      \end{tikzpicture}
    }
    \caption{}
    \label{fig:nbrs-example-reset}
  \end{subfigure}
  \caption{Non-Deterministic Bigraph Example. (a) Initial bigraph. (b)
    $A_{\mathit{send}}$ -- Attempt send, fails (orange) or succeeds (green).
    (c) $A_{\mathit{wait}}$ -- do nothing. (d) $A_{\mathit{reset}}$ --
    Get ready to send again.}
  \label{fig:nbrs-example}
\end{figure}
As an example, consider the model in \cref{fig:nbrs-example} representing
another simple WSN. In this case,   data can be sent between the
sensor (\textsf{S}) and base-station (\textsf{BS}), and  there is   a
non-deterministic \emph{choice}   whether the sensor should send data or wait.

\begin{figure}
  \centering
  \resizebox{0.5\linewidth}{!}{
  \begin{tikzpicture}
    \begin{scope}[local bounding box=init]
        \node[draw, minimum height=10] (bs) {};
        \node[lbl conc, anchor=south] at (bs.north) (bs_lbl) {$v_0$};
        \node[draw,circle] at ($(bs.east) + (0.5,0)$) (n1) {};
        \node[lbl conc, anchor=south west] at (n1.north east) (n1_lbl) {$v_1$};
        \draw[big edge] (bs.east) to[out=0, in=180] (n1);
        \node[big region, fit=(bs)(bs_lbl)(n1)(n1_lbl)] (r1) {};
\end{scope}
    \begin{scope}[shift={(2,1.5)}, local bounding box=sf]
        \node[draw, minimum height=10, fill=orange] (bs) {};
        \node[lbl conc, anchor=south] at (bs.north) (bs_lbl) {$v_0$};
        \node[draw,circle, fill=orange] at ($(bs.east) + (0.5,0)$) (n1) {};
        \node[lbl conc, anchor=south west] at (n1.north east) (n1_lbl) {$v_1$};
        \draw[big edge] (bs.east) to[out=0, in=180] (n1);
        \node[big region, fit=(bs)(bs_lbl)(n1)(n1_lbl)] (r1) {};
\end{scope}
    \begin{scope}[shift={(2,-1.5)}, local bounding box=ss]
        \node[draw, minimum height=10, fill=green] (bs) {};
        \node[lbl conc, anchor=south] at (bs.north) (bs_lbl) {$v_0$};
        \node[draw,circle, fill=green] at ($(bs.east) + (0.5,0)$) (n1) {};
        \node[lbl conc, anchor=south west] at (n1.north east) (n1_lbl) {$v_1$};
        \draw[big edge] (bs.east) to[out=0, in=180] (n1);
        \node[big region, fit=(bs)(bs_lbl)(n1)(n1_lbl)] (r1) {};
\end{scope}

\draw[-latex] (init.north) to[out=90, in=180, looseness=5] node[left, align=center]
    {\tiny $A_{\mathit{wait}}$ \\ \tiny 1} (init.west);

\draw[-latex] (sf.west) to[out=180, in=90, looseness=1] node[xshift=-5, near start, left, align=center]
    {\tiny $A_{\mathit{reset}}$ \\ \tiny 1} ($(init.north) + (0.2,0)$);

\draw[-latex] (init.east) to[out=0, in=-90, looseness=1] node[near end, right, align=center]
    {\tiny $\frac{\wfail}{w_\mathit{succ} + \wfail}$} ($(sf.south) + (0,0)$);
    \draw[-latex] (init.east) to[out=0, in=90, looseness=1] node[near end, right, align=center] {\tiny $\frac{w_{\mathit{succ}}}{w_\mathit{succ} + \wfail}$} ($(ss.north) + (0,0)$);

    \node[] at ($(init.east) + (1.2,0)$) {\tiny $A_{\mathit{send}}$};
  \end{tikzpicture}
  }
  \caption{Markov Decision Process for the example of \cref{fig:nbrs-example}. State $g_0$ left, $g_1$ bottom and $g_{2}$ top.}
  \label{fig:nbrs-example-mdp}
\end{figure}
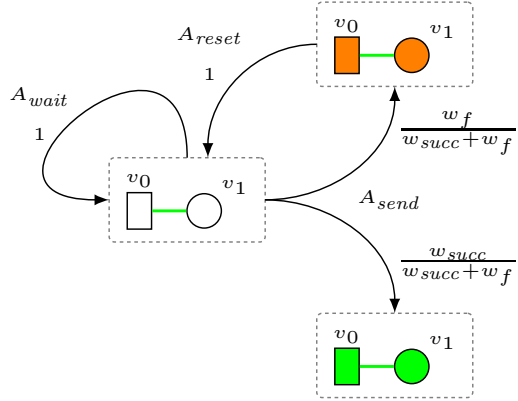

There are three actions: 
$A_{\textit{send}} = \{ \mathtt{send\_suc}, \mathtt{send\_fail} \}$,
$A_{\textit{wait}} = \{ \mathtt{wait} \}$, and
$A_{\textit{reset}} = \{ \mathtt{reset} \}$. The resulting Markov decision
process is shown in \cref{fig:nbrs-example-mdp}. 
From the initial state $g_{0}$
the system has a choice of two actions: $A_{\mathit{send}}$ or
$A_{\mathit{wait}}$. 
If the system \emph{chooses} to send then the distribution
of states is
$\distrf{g_{0}}{A_{\textit{send}}} = [ g_{1} \mapsto \frac{\wsuc}{\wsuc + \wfail}, g_{2} \mapsto \frac{\wfail}{\wsuc + \wfail} ]$,
while on a wait it is
$\distrf{g_{0}}{A_{\textit{wait}}} = [ 1 \mapsto g_{2} ]$. In the case
the send fails an additional action, $A_{\mathit{reset}}$ reinitialises the
state to allow another attempt.

\section{ Extended  BigraphER and further examples  }
\label{sec:examples}

We have extended BigraphER\footnote{Available online:
  \url{https://uog-bigraph.bitbucket.io/}}~\cite{DBLP:conf/cav/SevegnaniC16},
an open source toolkit for bigraphs, to support  standard, probabilistic,
stochastic, and action bigraphical reactive systems. BigraphER provides both a
language for describing bigraphs and reaction rules algebraically, and support
for simulating BRSs and generating transition systems.
In the following
we introduce the syntax of BigraphER by example, highlighting the new language constructs. A full reference is available~\cite{DBLP:conf/cav/SevegnaniC16}.

BigraphER uses modern (probabilistic) model checkers by exporting transition
systems to PRISM input format\footnote{These raw transition systems are often
  easily read by other model checkers.}. This enables quantitative model
checking for probabilistic (DTMC), stochastic (CTMC) and action (MDP) BRSs.
We present examples of each type and show selected BigraphER implementation details and results. 

\cref{sec:eval-virus} gives a probabilistic BRS  model of virus spread
through a computer network where firewalls are breached probabilistically. 

When
Stochastic Bigraphs were defined by Krivine
\etal~\cite{DBLP:journals/entcs/KrivineMT08} there was no tool support to
analyse their membrane budding model, instead requiring a manual translation to
a PRISM model. In \cref{sec:eval-stochastic} we revisit this model and for the first time, give a direct implementation in
bigraphs. 

Finally, in \cref{sec:eval-wsn}  we use an action BRS to model decisions  in wireless sensor
networks with mobile sinks: sensors decide whether to send immediately, or
wait in the hope the sink moves closer. 

Model files are available~\cite{modelFiles}.

\subsection{Probabilistic Bigraphs -- Virus Spread in a Network}
\label{sec:eval-virus}

Viruses spread through computer networks in a probabilistic manner. The   probability of infection may  differ between nodes, depending on the
effectiveness of firewalls. Although we focus on computer networks, similar
models could be adapted to, for example,  infection spread in populations.
This example is based on an existing PRISM
model~\cite{kwiatkowska.ea_ProbMobAmbs:2009}.

In an arbitrary computer network, with the infection starting at a
specified node, we wish to determine the likelihood of a virus infecting a
particular node. 
We use the following infection process.
Infected nodes attack, spreading the virus to uninfected neighbours. All (uninfected) neighbours are equally likely to be targets, \ie the rule has \emph{fixed} weight $w_{\textit{attack}}$. 
A node that has been attacked is either protected by a firewall with weight $w_{\textit{detect}}$, or become infected
with weight $w_{\textit{infect}}$. As we do not directly specify
the probabilities, the important measure is the \emph{ratio} of
$w_{\textit{detect}}$ to $w_{\textit{infect}}$. This differs from the example PRISM
model that adopts the more common approach of setting
$p_{\textit{detect}} = 1 - p_{\textit{infect}}$.

\begin{figure}
  \centering
  \begin{subfigure}{0.4\linewidth}
    \centering
    \resizebox{0.9\linewidth}{!}{
      \begin{tikzpicture}
        \begin{scope}[local bounding box=lhs]

          \begin{pgfonlayer}{foreground}
          \node[big site] (s1) {};
          \node[draw, circle, fill=black, inner sep=0.2, right=0.2 of s1,opacity=1] (e1) {};
          \end{pgfonlayer}
          \node[draw, circle, fill=vermilion, fit=(s1)(e1)] (n1) {};
          \node[lbl, anchor=south east] at (n1.north west) (n1_lbl) {I};

          \begin{pgfonlayer}{foreground}
          \node[big site, right=1.4 of s1] (s2) {};
          \node[draw, circle, fill=black, inner sep=0.2, left=0.2 of s2] (e2) {};
          \end{pgfonlayer}
          \node[draw, circle, black, fit=(s2)(e2)] (n2) {};
          \node[lbl, anchor=south east] at (n2.north west) (n2_lbl) {S};

          \node[big region, fit=(n1)(n1_lbl)] (r1) {};
          \node[big region, fit=(n2)(n2_lbl)] (r2) {};

          \draw[big edge] (e1.east) to[] (e2.west);
        \end{scope}

        \begin{scope}[shift={(4,0)}, local bounding box=rhs]
          \begin{pgfonlayer}{foreground}
          \node[big site] (s1) {};
          \node[draw, circle, fill=black, inner sep=0.2, right=0.2 of s1] (e1) {};
          \end{pgfonlayer}
          \node[draw, circle, fill=vermilion, fit=(s1)(e1)] (n1) {};
          \node[lbl, anchor=south east] at (n1.north west) (n1_lbl) {I};

          \begin{pgfonlayer}{foreground}
          \node[big site, right=1.4 of s1] (s2) {};
          \node[draw, circle, fill=black, inner sep=0.2, left=0.2 of s2] (e2) {};
          \end{pgfonlayer}
          \node[draw, circle, fill=ceruleanfrost, fit=(s2)(e2)] (n2) {};
          \node[lbl, anchor=south east] at (n2.north west) (n2_lbl) {A};

          \node[big region, fit=(n1)(n1_lbl)] (r1) {};
          \node[big region, fit=(n2)(n2_lbl)] (r2) {};

          \draw[big edge] (e1.east) to[] (e2.west);
        \end{scope}
        \node[] at ($(lhs.east)!0.5!(rhs.west)$) {$\rrulp{w_{\textit{attack}}}$};
      \end{tikzpicture}
    }
    \caption{\texttt{attack}}
    \label{fig:virus_spread_attack}
  \end{subfigure}
  \begin{subfigure}{0.25\linewidth}
    \centering
    \resizebox{\linewidth}{!}{
      \begin{tikzpicture}
        \begin{scope}[local bounding box=lhs]
          \begin{pgfonlayer}{foreground}
          \node[big site] (s1) {};
          \end{pgfonlayer}
          \node[draw, circle, fill=ceruleanfrost, fit=(s1)] (n1) {};
          \node[lbl, anchor=south east] at (n1.north west) (n1_lbl) {A};
          \node[big region, fit=(n1)(n1_lbl)] (r1) {};
        \end{scope}

        \begin{scope}[shift={(2.8,0)}, local bounding box=rhs]
          \begin{pgfonlayer}{foreground}
          \node[big site] (s1) {};
          \end{pgfonlayer}
          \node[draw, circle, fill=vermilion, fit=(s1)] (n1) {};
          \node[lbl, anchor=south east] at (n1.north west) (n1_lbl) {I};
          \node[big region, fit=(n1)(n1_lbl)] (r1) {};
        \end{scope}
        \node[] at ($(lhs.east)!0.5!(rhs.west)$) {$\rrulp{w_{\textit{infect}}}$} ;
      \end{tikzpicture}
    }
    \caption{\texttt{infect}}
    \label{fig:virus_spread_infect}
  \end{subfigure}
  \begin{subfigure}{0.25\linewidth}
    \centering
    \resizebox{\linewidth}{!}{
      \begin{tikzpicture}
        \begin{scope}[local bounding box=lhs]
          \begin{pgfonlayer}{foreground}
          \node[big site] (s1) {};
          \end{pgfonlayer}
          \node[draw, circle, fill=ceruleanfrost, fit=(s1)] (n1) {};
          \node[lbl, anchor=south east] at (n1.north west) (n1_lbl) {A};
          \node[big region, fit=(n1)(n1_lbl)] (r1) {};
        \end{scope}

        \begin{scope}[shift={(2.8,0)}, local bounding box=rhs]
          \begin{pgfonlayer}{foreground}
          \node[big site] (s1) {};
          \end{pgfonlayer}
          \node[draw, circle, fit=(s1)] (n1) {};
          \node[lbl, anchor=south east] at (n1.north west) (n1_lbl) {S};
          \node[big region, fit=(n1)(n1_lbl)] (r1) {};
        \end{scope}
        \node[] at ($(lhs.east)!0.5!(rhs.west)$) {$\rrulp{w_{\textit{detect}}}$} ;
      \end{tikzpicture}
    }
    \caption{\texttt{detect}}
    \label{fig:virus_spread_detect}
  \end{subfigure}
  \caption{Virus spread with  entities \textsf{S} = Safe, \textsf{I} =
    Infected, \textsf{A} = Attacked. (a) Attempt to attack an uninfected
    neighbour. (b) Firewall fails, node becomes infected. (c) Firewall succeeds.}
  \label{fig:virus_spread}
\end{figure}
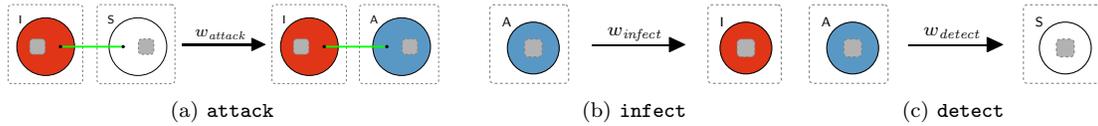

The bigraph model is in \cref{fig:virus_spread}. An \emph{infected} node
attacks a neighbouring (through the green link) \emph{safe} node (\textsf{S}, clear circle) with weight
$w_{\textit{attack}}$. An attacked node (\textsf{A}, indigo circle) enters the \emph{attacked} state and from there
has a $w_{\textit{infect}}$ weight of becoming infected (\textsf{I}, red circle), or
$w_{\textit{detect}}$ of the firewall breaking the infection attempt.

Unlike Safe-Infected-Recovered (SIR)
models~\cite{hethcote_MathematicsOfInfectiousDiseases:2000}, infected nodes
remain infected for the entire run, \ie there is no recovery, and there is no
resistance buildup, \ie if a firewall stops an infection the same node may still
be infected later.

\subsubsection{Syntax}

Probabilistic rules are defined in BigraphER by using keyword \texttt{react} and by specifying a weight (any
positive \texttt{float} expression) in the separator between the left-hand side and
right-hand side of a rule. For example, rule \texttt{infect} is defined by the code snippet in line \ref{code_infect_rule} of \cref{code_infect} with weight \texttt{w\_infect}. Entities are defined in lines 2-3 by using keyword \texttt{ctrl}. A PBRS is defined by construct \texttt{begin pbrs ... end} which also specifies the initial state, the state predicates, and the set of reaction rules (omitted).

\begin{bigrapher}[label={code_infect}, caption={Specifying probabilistic rules in BigraphER (snippet).}]
  # Defines entities with 0 arity (links)
  ctrl A = 0;
  ctrl I = 0;

  float w_infect = 5.0;

  # Probabilistic rule with weight w_infect
  react infect = A -[w_infect]-> I; @\label{code_infect_rule}@

  # Predicate
  # par(n,B) = B | B ... | B (n times)
  big all_infected = par(9, I); @\label{code_infect_pred}@

  begin pbrs
    @\dots@
  end
\end{bigrapher}

\subsubsection{Evaluation}

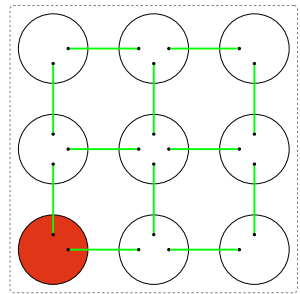
\begin{figure}
  \begin{subfigure}{0.6\linewidth}
  \centering
  {
  \setlength{\tabcolsep}{5pt}
  \begin{filecontents*}{virus-simple-grid-analysis.dat}
steps,five,ten,fifteen
50,0.04473748946082856,0.001220740282935578,9.586422951952695E-5
100,0.6088925114186892,0.08034169757184847,0.01181553609613111
250,0.9996148409192083,0.8879704306075762,0.5236567004985395
500,0.9997936243764995,0.997159386418325,0.9877394345923722
\end{filecontents*}
\pgfplotstabletypeset[
    col sep=comma,
    columns/steps/.style= {
      column type=c,
      column name=Max Steps,
      int trunc
    },
    columns/five/.style= {
      column type=c,
      column name={$5$},
      sci
    },
    columns/ten/.style= {
      column type=c,
      column name={$10$},
      sci
    },
    columns/fifteen/.style= {
      column type=c@{},
      column name={$15$},
      sci
    },
    columns={steps,five,ten,fifteen},
    every head row/.style={
        after row=\midrule, 
        before row={ & \multicolumn{3}{c}{$w_{\mathit{detect}}$} \\ }
    }] {virus-simple-grid-analysis.dat}
  }
  \caption{}
  \label{tab:virus_spread_simple_results}
  \end{subfigure}
  \begin{subfigure}{0.35\linewidth}
    \centering
    \resizebox{0.7\linewidth}{!}{
      \begin{tikzpicture}
        \begin{scope}[]
\coordinate (o1) at (0,0);
          \begin{pgfonlayer}{foreground}
          \node[draw, circle, fill=black, inner sep=0.2, right=0.2 of o1] (e1) {};
          \node[draw, circle, fill=black, inner sep=0.2, above=0.2 of o1] (e2) {};
          \node[draw, circle, fill=black, inner sep=0.2, below=0.2 of o1, opacity=0] (e3) {};
          \node[draw, circle, fill=black, inner sep=0.2, left=0.2 of o1,opacity = 0] (e4) {};
          \end{pgfonlayer}
          \node[draw, circle, fill=vermilion, fit=(e1)(e2)(e3)(e4)] (n1) {};

          \coordinate (o2) at (1.5,0);
          \node[draw, circle, fill=black, inner sep=0.2, right=0.2 of o2] (e5) {};
          \node[draw, circle, fill=black, inner sep=0.2, above=0.2 of o2] (e6) {};
          \node[draw, circle, fill=black, inner sep=0.2, below=0.2 of o2, opacity=0] (e7) {};
          \node[draw, circle, fill=black, inner sep=0.2, left=0.2 of o2,] (e8) {};
          \node[draw, circle, fit=(e5)(e6)(e7)(e8)] (n2) {};

          \coordinate (o3) at (3,0);
          \node[draw, circle, fill=black, inner sep=0.2, right=0.2 of o3, opacity=0] (e9) {};
          \node[draw, circle, fill=black, inner sep=0.2, above=0.2 of o3] (e10) {};
          \node[draw, circle, fill=black, inner sep=0.2, below=0.2 of o3, opacity=0] (e11) {};
          \node[draw, circle, fill=black, inner sep=0.2, left=0.2 of o3,] (e12) {};
          \node[draw, circle, fit=(e9)(e10)(e11)(e12)] (n3) {};

          \draw[big edge] (e1) to[] (e8);
          \draw[big edge] (e5) to[] (e12);
\coordinate (o4) at (0,1.5);
          \node[draw, circle, fill=black, inner sep=0.2, right=0.2 of o4] (e13) {};
          \node[draw, circle, fill=black, inner sep=0.2, above=0.2 of o4] (e14) {};
          \node[draw, circle, fill=black, inner sep=0.2, below=0.2 of o4] (e15) {};
          \node[draw, circle, fill=black, inner sep=0.2, left=0.2 of o4,opacity = 0] (e16) {};
          \node[draw, circle, fit=(e13)(e14)(e15)(e16)] (n4) {};

          \coordinate (o5) at (1.5,1.5);
          \node[draw, circle, fill=black, inner sep=0.2, right=0.2 of o5] (e17) {};
          \node[draw, circle, fill=black, inner sep=0.2, above=0.2 of o5] (e18) {};
          \node[draw, circle, fill=black, inner sep=0.2, below=0.2 of o5] (e19) {};
          \node[draw, circle, fill=black, inner sep=0.2, left=0.2 of o5] (e20) {};
          \node[draw, circle, fit=(e17)(e18)(e19)(e20)] (n5) {};

          \coordinate (o6) at (3,1.5);
          \node[draw, circle, fill=black, inner sep=0.2, right=0.2 of o6, opacity=0] (e21) {};
          \node[draw, circle, fill=black, inner sep=0.2, above=0.2 of o6] (e22) {};
          \node[draw, circle, fill=black, inner sep=0.2, below=0.2 of o6] (e23) {};
          \node[draw, circle, fill=black, inner sep=0.2, left=0.2 of o6] (e24) {};
          \node[draw, circle, fit=(e21)(e22)(e23)(e24)] (n6) {};
.
\draw[big edge] (e2) to[] (e15);
          \draw[big edge] (e6) to[] (e19);
          \draw[big edge] (e10) to[] (e23);

          \draw[big edge] (e13) to[] (e20);
          \draw[big edge] (e17) to[] (e24);

\coordinate (o7) at (0,3);
          \node[draw, circle, fill=black, inner sep=0.2, right=0.2 of o7] (e25) {};
          \node[draw, circle, fill=black, inner sep=0.2, above=0.2 of o7, opacity=0] (e26) {};
          \node[draw, circle, fill=black, inner sep=0.2, below=0.2 of o7] (e27) {};
          \node[draw, circle, fill=black, inner sep=0.2, left=0.2 of o7,opacity = 0] (e28) {};
          \node[draw, circle, fit=(e25)(e26)(e27)(e28)] (n7) {};

          \coordinate (o8) at (1.5,3);
          \node[draw, circle, fill=black, inner sep=0.2, right=0.2 of o8] (e29) {};
          \node[draw, circle, fill=black, inner sep=0.2, above=0.2 of o8, opacity=0] (e30) {};
          \node[draw, circle, fill=black, inner sep=0.2, below=0.2 of o8] (e31) {};
          \node[draw, circle, fill=black, inner sep=0.2, left=0.2 of o8] (e32) {};
          \node[draw, circle, fit=(e29)(e30)(e31)(e32)] (n8) {};

          \coordinate (o9) at (3,3);
          \node[draw, circle, fill=black, inner sep=0.2, right=0.2 of o9, opacity=0] (e33) {};
          \node[draw, circle, fill=black, inner sep=0.2, above=0.2 of o9, opacity=0] (e34) {};
          \node[draw, circle, fill=black, inner sep=0.2, below=0.2 of o9] (e35) {};
          \node[draw, circle, fill=black, inner sep=0.2, left=0.2 of o9] (e36) {};
          \node[draw, circle, fit=(e33)(e34)(e35)(e36)] (n9) {};

\draw[big edge] (e14) to[] (e27);
          \draw[big edge] (e18) to[] (e31);
          \draw[big edge] (e22) to[] (e35);

          \draw[big edge] (e25) to[] (e32);
          \draw[big edge] (e29) to[] (e36);

          \node[big region, fit=(n1)(n2)(n3)(n4)(n5)(n6)(n7)(n8)(n9)] (r1) {};
        \end{scope}
      \end{tikzpicture}
    }
    \caption{}
    \label{fig:virus_spread_simple_initial}
    \end{subfigure}
    \caption{(a) Probability of full infection in $n$ steps. (b) Initial topology}
\end{figure}

Consistent with the original PRISM model, we use a topology of nine
nodes connected in a square grid layout as in \cref{fig:virus_spread_simple_initial}.
Using BigraphER we export the full (probabilistic) transition system for further
analysis in PRISM. \Cref{tab:virus_spread_simple_results} shows the probability
that all nine nodes of the system are infected within the first $n$ timesteps
(reactions), \ie the Probabilistic Computation Tree Logic (PCTL)~\cite{hansson.jonsson_PCTL}
formula ${\mathcal P}_{=?}\left[\, {\bf F} ^{ \leq n } \mathtt{all\_infected} \, \right] $
where \texttt{all\_infected} is a predicate matching nine infected nodes as defined on line \ref{code_infect_pred} of \cref{code_infect}. In the BigraphER languages, predicates are defined as bigraphs with keyword \texttt{big}. Iterated operator \texttt{par} allows to concisely place $n$ bigraphs
side-by-side inside the same region\footnote{This corresponds to the iterated \emph{merge product} operator in the algebraic form of bigraphs.}.

We  vary the detection weight $w_{\mathit{detect}}$ of the firewalls ($5$, $10$, and $15$) and, as
expected, increasing $w_{\mathit{dectect}}$, \ie adding better firewalls,
  reduces the probability all nodes become infected within a given
time period.

\subsubsection{Model Extensibility -- Behaviour}

\begin{figure}
  \centering
  \begin{subfigure}{0.45\linewidth}
    \centering
    \resizebox{0.8\linewidth}{!}{
      \begin{tikzpicture}
        \begin{scope}[local bounding box=lhs]
          \begin{pgfonlayer}{foreground}
          \node[big site] (s1) {};
          \node[draw, right=0.2 of s1, fill=teagreen] (f1) {};
          \node[lbl, anchor=south] at (f1.north) (f1_lbl) {BasicFW};
          \end{pgfonlayer}
          \node[draw, circle, fill=ceruleanfrost, fit=(s1)(f1)(f1_lbl)] (n1) {};
          \node[lbl, anchor=south east] at (n1.north west) (n1_lbl) {A};
          \node[big region, fit=(n1)(n1_lbl)] (r1) {};
        \end{scope}

        \begin{scope}[shift={(4,0)}, local bounding box=rhs]
          \begin{pgfonlayer}{foreground}
          \node[big site] (s1) {};
          \node[draw, right=0.2 of s1, fill=teagreen] (f1) {};
          \node[lbl, anchor=south] at (f1.north) (f1_lbl) {BasicFW};
          \end{pgfonlayer}
          \node[draw, circle, fit=(s1)(f1)(f1_lbl)] (n1) {};
          \node[lbl, anchor=south east] at (n1.north west) (n1_lbl) {S};
          \node[big region, fit=(n1)(n1_lbl)] (r1) {};
        \end{scope}
        \node[] at ($(lhs.east)!0.5!(rhs.west)$) {$\rrulp{w_{detectBasic}}$} ;
      \end{tikzpicture}
    }
    \caption{\texttt{detect\_basic}}
    \label{fig:virus_spread_detect_new_basic}
  \end{subfigure}
  \begin{subfigure}{0.45\linewidth}
    \centering
    \resizebox{0.9\linewidth}{!}{
      \begin{tikzpicture}
        \begin{scope}[local bounding box=lhs]
          \begin{pgfonlayer}{foreground}
          \node[big site] (s1) {};
          \node[draw, right=0.2 of s1, fill=canary] (f1) {};
          \node[lbl, anchor=south] at (f1.north) (f1_lbl) {AdvFW};
          \end{pgfonlayer}
          \node[draw, circle, fill=ceruleanfrost, fit=(s1)(f1)(f1_lbl)] (n1) {};
          \node[lbl, anchor=south east] at (n1.north west) (n1_lbl) {A};
          \node[big region, fit=(n1)(n1_lbl)] (r1) {};
        \end{scope}

        \begin{scope}[shift={(4,0)}, local bounding box=rhs]
          \begin{pgfonlayer}{foreground}
          \node[big site] (s1) {};
          \node[draw, right=0.2 of s1, fill=canary] (f1) {};
          \node[lbl, anchor=south] at (f1.north) (f1_lbl) {AdvFW};
          \end{pgfonlayer}
          \node[draw, circle, fit=(s1)(f1)(f1_lbl)] (n1) {};
          \node[lbl, anchor=south east] at (n1.north west) (n1_lbl) {S};
          \node[big region, fit=(n1)(n1_lbl)] (r1) {};
        \end{scope}
        \node[] at ($(lhs.east)!0.5!(rhs.west)$) {$\rrulp{w_{detectAdv}}$} ;
      \end{tikzpicture}
    }
    \caption{\texttt{detect\_advanced}}
    \label{fig:virus_spread_detect_new_adv}
  \end{subfigure}
  \caption{Modified \texttt{detect} rule accounting for firewall types }
  \label{fig:virus_spread_detect_new}
\end{figure}
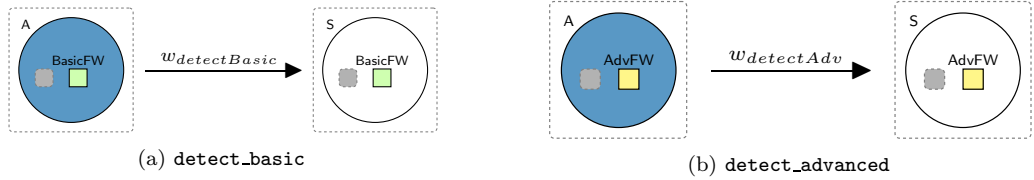

As shown, better firewalls reduce  the time to whole system infection, however
this can be prohibitively expensive in a deployed system. Alternatively, we may wish to improve the
firewalls on specific nodes. To test this, we add new entities
\textsf{BasicFW} (light green squares) and \textsf{AdvFW} (yellow squares) representing a basic and more advanced (and
expensive) firewall. We then split the \texttt{detect} rule into two rules, as
shown in \cref{fig:virus_spread_detect_new} to account for the type of
firewall. Because   the  sites  abstract away internal node structure,
the other rules remain the same. The only other change is to place the firewalls
in the starting topology -- in this case we place advanced firewalls in the
nodes of the middle row.

\begin{table}
  \centering
  \caption{Probability of full infection in $n$ steps when using 3 improved firewalls. In all cases $w_{\mathit{detectBasic}} = 5$.}
  {
  \setlength{\tabcolsep}{5pt}
  \begin{filecontents*}{virus-firewall-grid-analysis.dat}
steps,five,ten,fifteen
50,0.04473748946082856,0.011907459065537997,0.004696867050499343
100,0.6088925114186892,0.31240320934162225,0.16733811626896
250,0.9996148409192085,0.9879821943945459,0.9294881219538684
500,0.9997936243764993,0.99884394312444,0.9988192723887102
\end{filecontents*}
\pgfplotstabletypeset[
    col sep=comma,
    columns/steps/.style= {
      column type=c,
      column name=Max Steps,
      int trunc
    },
    columns/five/.style= {
      column type=c,
      column name={$5$},
      sci
    },
    columns/ten/.style= {
      column type=c,
      column name={$10$},
      sci
    },
    columns/fifteen/.style= {
      column type=c@{},
      column name={$15$},
      sci
    },
    every head row/.style={
        after row=\midrule, 
        before row={ & \multicolumn{3}{c}{$w_{\mathit{detectAdv}}$} \\ }
    },
    columns={steps,five,ten,fifteen},
    ] {virus-firewall-grid-analysis.dat}
  }
  \label{tab:virus_spread_simple_results_adv}
\end{table}

The results of re-running the analysis are in
\cref{tab:virus_spread_simple_results_adv}. As expected, when
$w_{\mathit{detectBasic}} = w_{\mathit{detectAdv}}$ the results match the
previous experiment. 
  Results like these can be used in a cost-benefit analysis,~e.g. when determining how many advanced firewalls to deploy.

\subsection{Stochastic Bigraphs -- Membrane Budding}
\label{sec:eval-stochastic}

Stochastic Bigraphs were first introduced by Krivine, Milner, and
Troina~\cite{DBLP:journals/entcs/KrivineMT08}, and allow BRS to form a CTMC (\cref{def:ctmc}) as an underlying transition system.
Instead of assigning \emph{weights} to reaction rule (as in PBRS), 
they assign a \emph{rate} to each reaction rule. 
Because these rates are defined over rules, not states, similar to  PBRS, a normalisation procedure is required to calculate a final exit rate based on the number of occurrences of each (applicable) rule.

The stochastic bigraphs paper presents a stochastic model of \emph{membrane budding}, which describes a biological mechanism for particles to move between cells.
In essence, coat proteins form on the surface of a membrane and, given enough coat proteins, a bud can form, accept some particles, and finally break off to carry particles out of a cell. Stochasticity comes from rates of coating, particle movement (into/out of the bud), and fission (breaking away).

No implementation of stochastic bigraphs was available when the example was developed,   and  analysis was given using a hand-coded PRISM model. As BigraphER now supports
stochastic bigraphs, we revisit this example to recreate their analysis
directly in bigraphs. 
We show   a snippet of the model here, and  refer the reader to the original source for full details of the
model \cite{DBLP:journals/entcs/KrivineMT08}.

We focus on the rule \texttt{coat} (rule 2 \cite{DBLP:journals/entcs/KrivineMT08}) that allows free \textsf{Coat} proteins to form on a \textsf{Bud}. The rule is shown in \cref{fig:stochastic_original}. Free \textsf{Coat} proteins are distinguished from those already forming on a bud by the use of a closed link. The site abstracts over the particles within the \textsf{Bud}.

The rate $rc$ determines how quickly \textsf{Coat}s form on the bud. Although it looks constant, because of the normalisation procedure, it in fact depends on the number of free coats, \ie the number of occurrences increases and, in turn, so does the rate the rule applies. This is similar to the example in \cref{sec:example_pbrs} where the probability \emph{any} sensor failed was much higher than the probability a specific single sensor failed.

\begin{figure}
  \centering
\resizebox{0.5\linewidth}{!}{
      \begin{tikzpicture}
        \begin{scope}
          \begin{scope}[local bounding box=lhs]
            \node[big site] (s1) {};
            \node[circle, draw, fit=(s1)] (bud) {};
            \node[lbl, anchor=north west] at (bud.south) (bud_lbl) {Bud};

            \node[circle, draw, right=0.3 of bud] (coatn) {};
            \node[lbl, anchor=west] at (coatn.east) (coatn_lbl) {Coat};
            \node[big region, fit=(bud)(bud_lbl)(coatn)(coatn_lbl)] (r1) {};
            \node[above=0.15 of r1] (x) {$x$};

            \draw[big edge] (bud.north) to[out=100, in=-90] (x.south);
            \draw[big edgec] (coatn.west) to[out=180, in=0] ($(coatn.west) + (-0.1,0.00)$);
          \end{scope}
          \begin{scope}[shift={(2.8,0)}, local bounding box=rhs]
            \node[big site] (s1) {};
            \node[circle, draw, fit=(s1)] (bud) {};
            \node[lbl, anchor=north west] at (bud.south) (bud_lbl) {Bud};

            \node[circle, draw, right=0.3 of bud] (coatn) {};
            \node[lbl, anchor=west] at (coatn.east) (coatn_lbl) {Coat};
            \node[big region, fit=(bud)(bud_lbl)(coatn)(coatn_lbl)] (r1) {};
            \node[above=0.15 of r1] (x) {$x$};

            \coordinate (h1) at ($(bud.north) + (0.51,0.1)$);
            \draw[big edge] (bud.north) to[out=100, in=-90] (h1);
            \draw[big edge] (h1) to[out=100, in=-90] (x);
            \draw[big edge] (coatn.west) to[out=180, in=-90] (h1);
          \end{scope}
          \node[yshift=-6] at ($(lhs.east)!0.5!(rhs.west)$) {$\rrulp{rc}$};
        \end{scope}
      \end{tikzpicture}
    }
    \caption{\texttt{coat} (recreated from \cite{DBLP:journals/entcs/KrivineMT08})}
    \label{fig:stochastic_original}
\end{figure}

Stochastic rules are specified in a similar way to probabilistic rules in
BigraphER by placing a float expression between the left and right-hand side of
a rule. For example, the rule \texttt{coat} is as follows.
\begin{bigrapher}[caption={Specifying stochastic rules in BigraphER (snippet).}, label={code_buds}]
  ctrl Bud = 1; # Arity 1
  ctrl Coat = 1;
  ctrl Gate = 1;
  ctrl Particle = 0; # No links

  # Coating rate
  float rc = 1.0;

  react coat =
    Bud{x}.(id | Gate{z}) | /y Coat{y}
    -[rc]->
    Bud{x}.(id | Gate{z}) | Coat{x};

  # Predicate used for plotting. 
  # Determines the number of particles that have been transferred from a membrane when the bud breaks free
  # par(n, b) = b | ... | b (n times)
  fun big particles(n) = Bud{x}.(par(n, Particle)); @\label{code_particles_pred}@

  begin sbrs
    @\dots@
  end
\end{bigrapher}
Similarly, the initial state and the reaction rules of an SBRS are defined by construct \texttt{begin sbrs ... end}.

\begin{figure}
  \centering
  \resizebox{0.8\linewidth}{!}{
    \begin{tikzpicture}
      \begin{scope}
        \begin{scope}[local bounding box=lhs]
          \node[big site] (s1) {};
          \node[circle, draw, fit=(s1)] (bud) {};
          \node[lbl, anchor=north west] at (bud.south) (bud_lbl) {Bud};
          \node[circle, draw] at ($(bud.east) + (0.3,0.2)$) (coat) {};
          \node[lbl, anchor=west] at (coat.east) (coat_lbl) {Coats($c$)};

          \node[circle, draw] at ($(bud.east) + (0.3,-0.3)$) (coatn) {};
          \node[lbl, anchor=west] at (coatn.east) (coatn_lbl)
          {Fcoats($c_{\mathit{max}} - c$)};
          \node[big region, fit=(bud)(bud_lbl)(coat)(coat_lbl)(coatn)(coatn_lbl)] (r1) {};
          \node[above=0.15 of r1, xshift=-25] (x) {$x$};

          \coordinate (h1) at ($(bud.north) + (0.1,0.2)$);
          \draw[big edge] (bud.north) to[out=100, in=-90] (h1);
          \draw[big edge] (h1) to[out=100, in=-90] (x.south);
          \draw[big edge] (coat.west) to[out=180, in=-90] (h1);
        \end{scope}
        \begin{scope}[shift={(5,0)}, local bounding box=rhs]
          \node[big site] (s1) {};
          \node[circle, draw, fit=(s1)] (bud) {};
          \node[lbl, anchor=north west] at (bud.south) (bud_lbl) {Bud};
          \node[circle, draw] at ($(bud.east) + (0.3,0.2)$) (coat) {};
          \node[lbl, anchor=west] at (coat.east) (coat_lbl) {Coats($c + 1$)};

          \node[circle, draw] at ($(bud.east) + (0.3,-0.3)$) (coatn) {};
          \node[lbl, anchor=west] at (coatn.east) (coatn_lbl)
          {Fcoats($c_{\mathit{max}} - c - 1$)};
          \node[big region, fit=(bud)(bud_lbl)(coat)(coat_lbl)(coatn)(coatn_lbl)] (r1) {};
          \node[above=0.15 of r1, xshift=-25] (x) {$x$};

          \coordinate (h1) at ($(bud.north) + (0.1,0.2)$);
          \draw[big edge] (bud.north) to[out=100, in=-90] (h1);
          \draw[big edge] (h1) to[out=100, in=-90] (x.south);
          \draw[big edge] (coat.west) to[out=180, in=-90] (h1);
        \end{scope}
        \node[yshift=-6] at ($(lhs.east)!0.5!(rhs.west)$) {$\rrulp{rc \cdot (c_{\mathit{max}} - c)}$};
      \end{scope}
    \end{tikzpicture}
  }
  \caption{\texttt{coat} -- population model.}
  \label{fig:stochastic-population}
\end{figure}
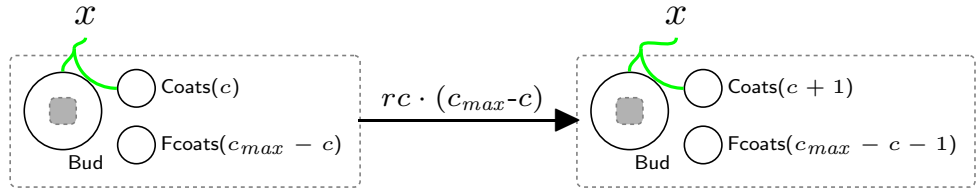

For large models \eg those with 50 free \textsf{Coat}  entities, it
is time consuming to count occurrences. To overcome this, we use a population model (a {\em counter abstraction}) that
groups free coats \etc into a single entity \textsf{Coats} representing the number bound, and \textsf{Fcoats} representing the number that are still free.
This allows \texttt{coat} to be alternatively written as
shown in \cref{fig:stochastic-population}.
Entities such as $\mathsf{Coats}(c)$ are \emph{parameterised entities} that can be seen as defining a family of \textsf{Coats} entities, one for each possible value of $c$. To calculate the number of free coats (\textsf{Fcoats}), we use the constant $c_\mathit{max}$ -- the total number of coat proteins (free or bound); in our case 50. \textsf{Coats} are defined in the BigraphER language by the \texttt{fun ctrl} keywords.
Similarly we augment the other rules
to, for example, count the number of \textsf{Protein}s in the \textsf{Bud}.
In this case we cannot rely on the occurrence count to determine the correct rate, so instead we explicitly scale the rate based on the number of remaining free coats.

For analysis, we have defined a family of bigraph predicates $\mathtt{particles(}n\mathtt{)}$, with $0\leq n \leq 40$, that match a \textsf{Bud} with exactly $n$ particles inside, allowing the probability that we end with $n$ particles in a bud (that has broken off) to be determined through the PCTL formula: ${\mathcal P}_{=?}\left[\, {\bf F}\, \mathtt{particles(}n\mathtt{)} \, \right] $. This is defined in line \ref{code_particles_pred} of \cref{code_buds} by using the \texttt{fun} keyword.

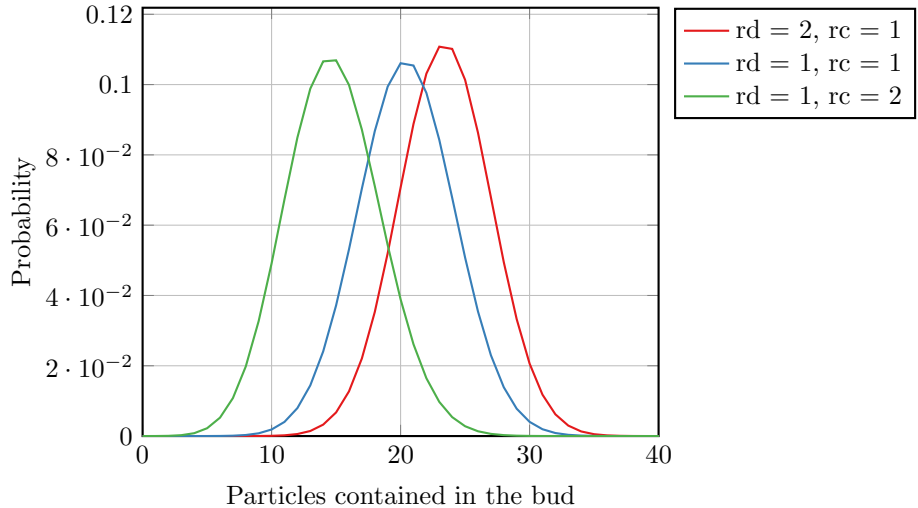
\begin{figure}
  \centering
  \resizebox{0.75\linewidth}{!}{
  \begin{tikzpicture}
    \pgfplotsset{
      cycle list/Set1-5,
    }
    \begin{axis}[
      xlabel=Particles contained in the bud,
      ylabel=Probability,
      grid=major,
thick,
      xmin=0, xmax=40,
      ymin=0,
      legend pos=outer north east,
      ylabel style={yshift=10},
      ]

      \addplot table [x expr=\coordindex, y = particles] {particles_1_1_2.dat};

      \addplot table [x expr=\coordindex, y = particles] {particles_1_1_1.dat};

      \addplot table [x expr=\coordindex, y = particles] {particles_1_2_1.dat};

      \legend{{rd = 2, rc = 1}, {rd = 1, rc = 1}, {rd = 1, rc = 2}};
    \end{axis}
  \end{tikzpicture}
  }
  \caption{Reproducing ``Particles in the formed bud'' figure from
    \cite{DBLP:journals/entcs/KrivineMT08} directly in bigraphs.}
  \label{fig:stochastic-particles-formed}
\end{figure}

We exported the BigraphER   model to PRISM,  where  we successfully reproduced the
``Particles in the formed bud'' results from the appendix of stochastic
bigraphs~\cite{DBLP:journals/entcs/KrivineMT08}, as shown in 
\cref{fig:stochastic-particles-formed}.
This figure shows the number of particles that are in the bud after fission, where $rd$ is the rate of diffusion of particles between the membrane and the bud (rule 3 \cite{DBLP:journals/entcs/KrivineMT08}). As expected, increasing the diffusion rate increases the expected number of particles in the formed bud.
The rate of fission depends on the number of coat proteins (more coat proteins implying higher fission rates), and hence for $rc = 2$ the overall effect is to have less expected particles in the bud.

\subsection{Action Bigraphs -- Mobile Sinks in Wireless Sensor Networks}
\label{sec:eval-wsn}

  We
use action bigraphs to model a well known decision problem in wireless sensor networks~\cite{boloni.ea_ShouldISendNowOrLater:2008}.
Traditionally, WSNs use multi-hop communication to move data between sensors and
fixed sink (base-station). An alternative approach is for a \emph{moving}
sink to collect data directly from the sensors. Such an approach can be used, for example,
when robots obtain information as they move through a space.

Given limited battery and memory capabilities of a sensor, when the sink moves
into range a decision must be made: should the sensor send immediately, with
high transmit power, or should it wait in the hope the sink moves closer,
risking losing data by exhausting memory if the sink moves out of range before
transmission can occur? Such decisions can be modelled as \emph{actions}.

\begin{figure}
  \centering
  \begin{subfigure}[b]{0.45\linewidth}
    \centering
    \resizebox{0.4\linewidth}{!}{
      \begin{tikzpicture}
        \begin{scope}[]
          \node[sink] (sink) {};
          \node[big site, minimum size=15, below=0.05 of sink, opacity=0] (s1) {};
          \node[dist, fit=(sink), minimum size=40] (d1) {};
          \node[big site, minimum size=15, below=0.3 of s1, opacity=0] (s2) {};
          \node[dist, fit=(d1), minimum size=80] (d3) {};
          \node[node, right=1.4 of sink] (n1) {$0$};
          \node[node, left=1.4 of s1] (n2) {$0$};
          \node[dist, draw, fit=(d1), minimum size=140] (d3) {};
          \node[big region, fit=(d3)] (r1) {};
        \end{scope}
      \end{tikzpicture}
    }
    \caption{Initial Bigraph}
    \label{fig:snol-init}
  \end{subfigure}
  \begin{subfigure}[b]{0.45\linewidth}
    \centering
    \resizebox{0.9\linewidth}{!}{
    \begin{tikzpicture}
        \begin{scope}[]
          \node[sink, minimum size=20] (sink) {};
          \node[lbl, anchor=north] at (sink.south) (sink_lbl) {Mobile Sink};
        \end{scope}
        \begin{scope}[shift={(3,0)}]
          \node[node] (n1) {$b$};
          \node[lbl, anchor=north] at (n1.south) (lbl) {Node containing buffer with $b$ elements};
        \end{scope}
        \begin{scope}[shift={(1.5,-1.2)}]
          \node[sink, minimum size=5, opacity=0] (sink) {};
          \node[dist, fit=(sink)] (c1) {};
          \node[lbl, anchor=north] at (c1.south) (lbl) {Distance Boundary};
        \end{scope}
    \end{tikzpicture}
  }
  \caption{Entities}
  \label{fig:snol-entities}
\end{subfigure}
\caption {Send now or later: Model elements}
\end{figure}
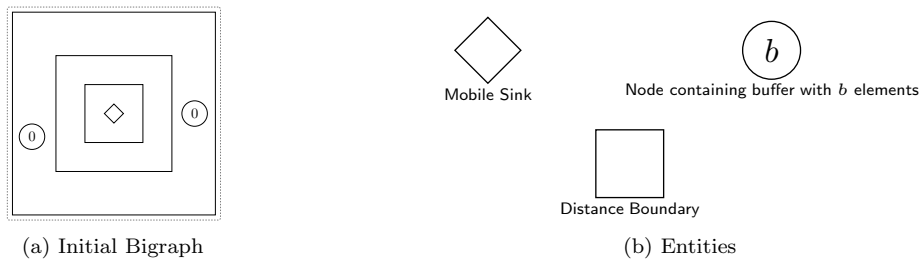

Model entities are shown in \cref{fig:snol-init}. There are three types:
a (unique) \textsf{Sink}; a sensor \textsf{Node} that includes a \textsf{Buffer}
with the number of slots filled; and \textsf{Distance} boundaries abstractly
representing how close a sensor is to the sink, \eg close, mid-range, or far.
\text{Node}s outwith the last \textsf{Distance} entity are considered
out-of-range and cannot send.

The model updates in two phases: \emph{movement}, where \textsf{Node}s
(possibly) move between \textsf{Distance} entities, and \emph{act}, where
\textsf{Node}s non-deterministically \emph{decide} whether to send or wait and
receive data. Phases are scheduled in a round-robin fashion with each
\textsf{Node} taking an action before moving onto the next phase.

\begin{figure}
  \centering
  \begin{subfigure}[b]{0.6\linewidth}
    \centering
    \resizebox{0.7\linewidth}{!}{
      \begin{tikzpicture}
        \begin{scope}[local bounding box=lhs]
          \node[big site] (s1) {};
          \node[big site, right=0.3 of s1] (s2) {};
          \node[node, fit=(s2)] (n1) {};

          \node[big site, right=0.65 of s2, opacity=0] (s3) {};
          \node[node, fit=(s3), opacity=0] (n2) {};

          \node[dist, fit=(s1)(n1)] (d1) {};
          \node[dist, minimum size=80, fit=(d1)] (d2) {};
          \node[big region, fit=(d2)] (r1) {};
        \end{scope}
        \begin{scope}[shift={(4.4,0)}, local bounding box=rhs]
          \node[big site] (s1) {};
          \node[big site, right=0.3 of s1, opacity=0] (s2) {};
          \node[node, fit=(s2), opacity=0] (n1) {};

          \node[big site, right=0.65 of s2, opacity=1] (s3) {};
          \node[node, fit=(s3), opacity=1] (n2) {};

          \node[dist, fit=(s1)(n1)] (d1) {};
          \node[dist, minimum size=80, fit=(d1)] (d2) {};
          \node[big region, fit=(d2)] (r1) {};
        \end{scope}
        \node[] at ($(lhs.east)!0.5!(rhs.west)$) {$\rrulp{w_{\mathit{move}}}$} ;
        \begin{scope}[shift={(0,-4)}, local bounding box=lhs2]
          \node[big site] (s1) {};
          \node[big site, right=0.3 of s1, opacity=0] (s2) {};
          \node[node, fit=(s2), opacity=0] (n1) {};

          \node[big site, right=0.65 of s2, opacity=1] (s3) {};
          \node[node, fit=(s3), opacity=1] (n2) {};

          \node[dist, fit=(s1)(n1)] (d1) {};
          \node[dist, minimum size=80, fit=(d1)] (d2) {};
          \node[big region, fit=(d2)] (r1) {};
        \end{scope}
        \begin{scope}[shift={(4.4,-4)}, local bounding box=rhs2]
          \node[big site] (s1) {};
          \node[big site, right=0.3 of s1, opacity=1] (s2) {};
          \node[node, fit=(s2), opacity=1] (n1) {};

          \node[big site, right=0.65 of s2, opacity=0] (s3) {};
          \node[node, fit=(s3), opacity=0] (n2) {};

          \node[dist, fit=(s1)(n1)] (d1) {};
          \node[dist, minimum size=80, fit=(d1)] (d2) {};
          \node[big region, fit=(d2)] (r1) {};
        \end{scope}
        \node[] at ($(lhs2.east)!0.5!(rhs2.west)$) {$\rrulp{w_{\mathit{move}}}$} ;
      \end{tikzpicture}
    }
    \caption{$A_{\mathtt{move}}$}
    \label{fig:snol-move}
  \end{subfigure}
  \begin{subfigure}[b]{0.35\linewidth}
    \centering
    \resizebox{0.7\linewidth}{!}{
      \begin{tikzpicture}
        \begin{scope}[local bounding box=lhs]
          \node[node, opacity=0] (n_invisible) {$b + 1$};
          \node[node] (n1) {$b$};
          \node[big region, fit=(n1)(n_invisible)] (r1) {};
        \end{scope}
        \begin{scope}[shift={(2.5,0)}, local bounding box=rhs]
          \node[node] (n1) {$b + 1$};
          \node[big region, fit=(n1)] (r1) {};
        \end{scope}
        \node[] at ($(lhs.east)!0.5!(rhs.west)$) {$\rrulp{\mathit{w_{suc}}}$} ;
        \begin{scope}[shift={(0,-2)}, local bounding box=lhs2]
          \node[node, opacity=0] (n_invisible) {$b + 1$};
          \node[node] (n1) {$b$};
          \node[big region, fit=(n1)(n_invisible)] (r1) {};
        \end{scope}
        \begin{scope}[shift={(2.5,-2)}, local bounding box=rhs2]
          \node[node] (n1) {$b$};
          \node[node, opacity=0] (n2) {$b + 1$};
          \node[big region, fit=(n2)] (r1) {};
        \end{scope}
        \node[] at ($(lhs2.east)!0.5!(rhs2.west)$) {$\rrulp{\mathit{w_{fail}}}$} ;
      \end{tikzpicture}
    }
    \caption{$A_{\mathtt{receive}}$ \scriptsize{$(b \le b_{max})$}}
    \label{fig:snol-receive}
  \end{subfigure}
  
  \begin{subfigure}[b]{0.45\linewidth}
    \centering
    \resizebox{0.7\linewidth}{!}{
      \begin{tikzpicture}
        \begin{scope}[local bounding box=lhs]
          \node[node] (n1) {$b_{max}$};
          \node[big region, fit=(n1)] (r1) {};
        \end{scope}
        \begin{scope}[shift={(2.5,0)}, local bounding box=rhs]
          \node[node] (n1) {$b_{max}$};
          \node[big region, fit=(n1)] (r1) {};
        \end{scope}
        \node[] at ($(lhs.east)!0.5!(rhs.west)$) {$\rrulp{\mathit{w_{suc}}}$} ;
      \end{tikzpicture}
    }
    \caption{$A_{\mathtt{receive\_{full}}}$; \scriptsize{Cost = 1}}
    \label{fig:snol-receive-max}
  \end{subfigure}
  \begin{subfigure}[b]{0.45\linewidth}
    \centering
    \resizebox{0.7\linewidth}{!}{
      \begin{tikzpicture}
        \begin{scope}[local bounding box=lhs]
          \node[sink, minimum size=20] (sink) {};
          \node[big site, minimum size=10, right=0.3 of sink] (s1) {};
          \node[node, right=0.3 of s1] (n1) {$b$};
          \node[dist, minimum size=60, fit=(n1)(s1)(sink)] (d1) {};
          \node[big region, fit=(d1)] (r1) {};
        \end{scope}
        \begin{scope}[shift={(3.5,0)}, local bounding box=rhs]
          \node[sink, minimum size=20] (sink) {};
          \node[big site, minimum size=10, right=0.3 of sink] (s1) {};
          \node[node, right=0.3 of s1] (n1) {$0$};
          \node[dist, minimum size=60, fit=(n1)(s1)(sink)] (d1) {};
          \node[big region, fit=(d1)] (r1) {};
        \end{scope}
        \node[] at ($(lhs.east)!0.5!(rhs.west)$) {$\rrulp{1}$} ;
      \end{tikzpicture}
    }
    \caption{$A_{\mathtt{send\_close}}$; \scriptsize{Cost = 1}}
    \label{fig:snol-sendc}
  \end{subfigure}
  
  \begin{subfigure}[b]{0.45\linewidth}
    \centering
    \resizebox{0.7\linewidth}{!}{
      \begin{tikzpicture}
        \begin{scope}[local bounding box=lhs]
          \node[sink] (sink) {};
          \node[big site, minimum size=10, right=0.2 of sink] (s1) {};
          \node[dist, fit=(sink)(s1), minimum size=40] (d1) {};
          \node[node, right=0.85 of sink] (n1) {$b$};
          \node[dist, fit=(d1), minimum size=80] (d2) {};
          \node[big region, fit=(d2)] (r1) {};
        \end{scope}
        \begin{scope}[shift={(4,0)}, local bounding box=rhs]
          \node[sink] (sink) {};
          \node[big site, minimum size=10, right=0.2 of sink] (s1) {};
          \node[dist, fit=(sink)(s1), minimum size=40] (d1) {};
          \node[node, right=0.85 of sink] (n1) {$0$};
          \node[dist, fit=(d1), minimum size=80] (d2) {};
          \node[big region, fit=(d2)] (r1) {};
        \end{scope}
        \node[] at ($(lhs.east)!0.5!(rhs.west)$) {$\rrulp{1}$} ;
      \end{tikzpicture}
    }
    \caption{$A_{\mathtt{send\_{mid}}}$; \scriptsize{Cost = 2}}
    \label{fig:snol-sendm}
  \end{subfigure}
  \begin{subfigure}[b]{0.5\linewidth}
    \centering
    \resizebox{0.7\linewidth}{!}{
      \begin{tikzpicture}
        \begin{scope}[local bounding box=lhs]
          \node[sink] (sink) {};
          \node[big site, minimum size=10, right=0.2 of sink] (s1) {};
          \node[dist, fit=(sink)(s1), minimum size=40] (d1) {};
          \node[big site, minimum size=10, right=0.4 of s1] (s1) {};
          \node[dist, fit=(d1), minimum size=80] (d2) {};
          \node[node, right=1.8 of sink] (n1) {$b$};
          \node[dist, fit=(d2), minimum size=140] (d3) {};

          \node[big region, fit=(d3)] (r1) {};
        \end{scope}
        \begin{scope}[shift={(6,0)}, local bounding box=rhs]
          \node[sink] (sink) {};
          \node[big site, minimum size=10, right=0.2 of sink] (s1) {};
          \node[dist, fit=(sink)(s1), minimum size=40] (d1) {};
          \node[big site, minimum size=10, right=0.4 of s1] (s1) {};
          \node[dist, fit=(d1), minimum size=80] (d2) {};
          \node[node, right=1.8 of sink] (n1) {$0$};
          \node[dist, fit=(d2), minimum size=140] (d3) {};
          \node[big region, fit=(d3)] (r1) {};
        \end{scope}
        \node[] at ($(lhs.east)!0.5!(rhs.west)$) {$\rrulp{1}$} ;
      \end{tikzpicture}
    }
    \caption{$A_{\mathtt{send\_{far}}}$; \scriptsize{Cost = 3}}
    \label{fig:snol-sendf}
  \end{subfigure}
  \caption {Send now or later: Actions, associated reaction rules, and costs}
  \label{fig:snol-react}
\end{figure}
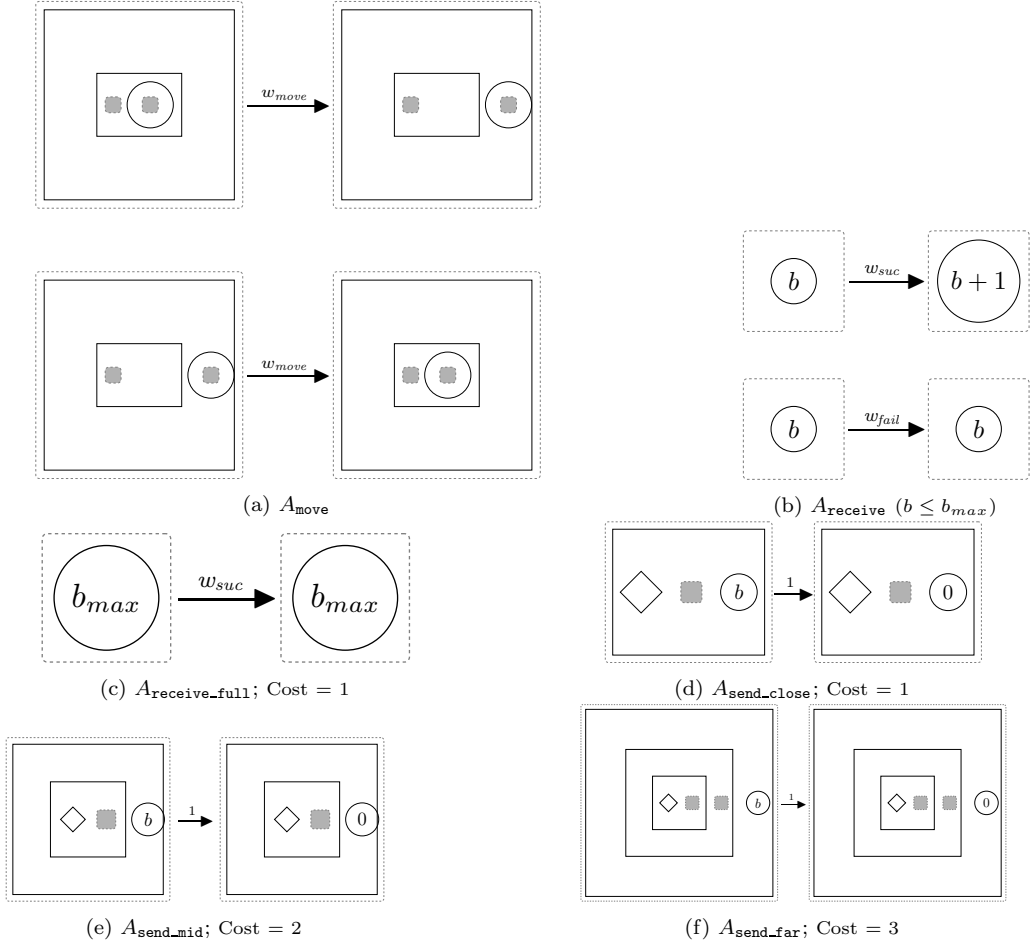

\textsf{Sink} movement is modelled by moving \textsf{Node}s between
\textsf{Distance} boundaries, as shown in the reaction rule of
\cref{fig:snol-move}. This models movement relative to the \textsf{Sink}, \ie a
\textsf{Node} moving to a closer \textsf{Distance} boundary implies a physical
\textsf{Sink} move towards the \textsf{Node}. An additional rule (unshown) moves
\textsf{Node}s in/out of the outermost \textsf{Distance} boundary.

Non-determinism in the operation phase comes from the decision to send existing data, through one of the $A_{\mathtt{send\_close}}$, $A_{\mathtt{send\_mid}}$, $A_{\mathtt{send\_far}}$ actions, or to receive new data through one of the $A_{\mathtt{receive}}$, $A_{\mathtt{receive\_full}}$ actions.
Using multiple actions to model send/receive allows multiple action reward structures to be used.
$A_{\mathtt{receive}}$ is shown in \cref{fig:snol-receive} and consists of two reaction rules representing receipt or failure of new data with weight $w_{\textit{receive}}$.
When the buffer is full ($b = b_{\textit{max}}$), 
$A_{\mathtt{receive}}$ is no longer applicable, and instead $A_{\mathtt{receive\_full}}$ (\cref{fig:snol-receive-max}) represents a dropped sample and unit cost is incurred.

Actions/Rules for sending data in
\cref{fig:snol-sendc,fig:snol-sendm,fig:snol-sendf}. In each case the buffer is fully emptied
and a cost is incurred proportional to the distance from the sensor. We assume
the cost is constant regardless of how much data is sent \ie data always fits in
a single radio packet. A more complex model could account for this through 
additional actions. Sending data is always successful but out-of-range sensors
are unable to send any data, \ie we do not fall back to hop-to-hop
communication, and must always do a receive.

\subsubsection{Syntax}

To specify ABRSs we extend the BigraphER language to include an explicit
\texttt{actions} declaration within \texttt{begin abrs ... end} blocks, as shown in \cref{lst:actions} (line 20). Each action
consists of an identifier, \eg \texttt{receive}, \texttt{receive\_full}, followed by a set of reaction rules, specified by their identifiers.
We may optionally assign a reward to each action, \eg \texttt{receive\_full[1]}
has reward/cost~1. 

\begin{bigrapher}[float,caption={Specifying actions BigraphER (snippet).}, label=lst:actions]
  ctrl Sensor = 0;
  fun ctrl Buffer(x) = 0;
  fun ctrl Iden(i) = 0;

  float w_suc = 5.0;
  float w_fail = 1.0;

  fun react receive(x,i) =
    Sensor.(Buffer(x) | Iden(i))
    -[w_suc]->
    Sensor.(Buffer(x + 1) | Iden(i))

  fun react receive_fail(x,i) =
    Sensor.(Buffer(x) | Iden(i))
    -[w_fail]->
    Sensor.(Buffer(x) | Iden(i))

  begin abrs
    @\dots@
    actions = [
      # Action[cost] = { rules }
      receive = {receive(x,i), receive_fail(x,i)},
      receive_full[1] = {receive(bmax,i)}, 
    ];
  end
\end{bigrapher}

\subsubsection{Model Analysis}

The model is analysed using PRISM~\cite{DBLP:conf/cav/KwiatkowskaNP11} by
exporting the MDP transition system from BigraphER. PRISM does not support
importing action rewards from a transition system, so we encode action
rewards through state rewards which are fully supported. To do this we add additional entities (unshown) to the model when actions are taken, \eg a \textsf{SendClose} entity if a sensor sends when the sink is close. These entities can be matched using state predicates to increase the cost. Importantly, this is an implementation detail to overcome tool support, and does not invalidate the theory.

We assume a single sink, two sensors, and 4 distance
boundaries: close-range, mid-range, far-range, and out-of-range. Both sensors
start in far-range of the sink.

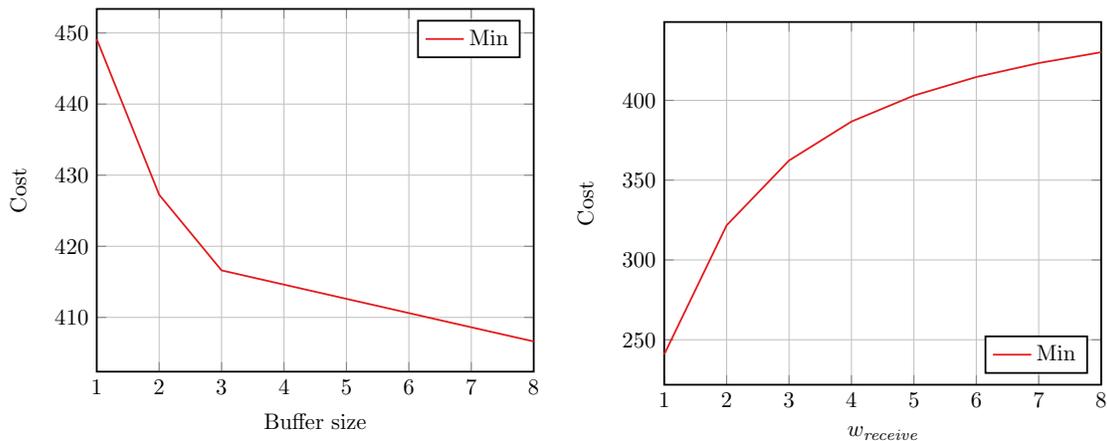
\begin{figure}
  \centering
  \begin{subfigure}{0.45\linewidth}
    \centering
    \resizebox{\linewidth}{!}{
      \begin{tikzpicture}
        \pgfplotsset{
          cycle list/Set1-5,
        }
        \begin{axis}[
          xlabel=Buffer size,
          ylabel=Cost,
          grid=major,
thick,
          xmin=1, xmax=8,
legend pos=north east,
]

          \addplot table [x = BufferSize, y = MinCost] {buffer-rewards.dat};

          \legend{Min};
        \end{axis}
      \end{tikzpicture}
    }
    \caption{Effect of increasing max buffer size on cost; $w_{\textit{receive}} = 6$.}
    \label{fig:buffer-cost}
  \end{subfigure}
  \begin{subfigure}{0.45\linewidth}
    \centering
    \resizebox{\linewidth}{!}{
      \begin{tikzpicture}
        \pgfplotsset{
          cycle list/Set1-5,
        }
        \begin{axis}[
          xlabel=$w_{\textit{receive}}$,
          ylabel=Cost,
          grid=major,
thick,
          xmin=1, xmax=8,
legend pos=south east,
]

          \addplot table [x = WeightRec, y = MinCost] {prec-rewards.dat};

          \legend{Min};
        \end{axis}
      \end{tikzpicture}
    }
    \caption{Effect of increasing $w_{\textit{receive}}$ on cost; Buffer size = 4}
    \label{fig:prec-cost}
  \end{subfigure}
  \caption{ Effects of buffer size and data receive weight on cost.}
  \label{fig:snol-params}
\end{figure}

First,  consider the effect of increasing the maximum buffer size on
the total cost, with the expected result being that an increased buffer size
should reduce the overall cost, since sensors can wait longer before deciding to
send data.
\Cref{fig:buffer-cost} shows how increasing the maximum buffer size effects the
minimum possible cost in the first 4000 transitions, using the PCTL formula
${\mathcal R}_{min=?}\left[\, {\bf C} ^{ \leq 4000 }\, \right]$.
As expected, the minimum cost reduces as the
buffer size increases as sensors can now delay sending for longer without
incurring a penalty. The relationship is non-linear and we can see that
increasing the buffer from 2 to 3 reading is much more beneficial than from 3 to
4.

Second,  consider  cost  reduction by  altering the probability that a sensor
receives data when not sending --  essentially reducing the sampling rate.
\Cref{fig:prec-cost} shows how changing $w_{\mathit{receive}}$ affects the
minimum possible cost for a fixed buffer size. As expected, increasing the
likelihood of receiving data increases overall costs, as the buffers are likely
to fill quickly and readings may be missed.

\paragraph{Extensibility} As with the virus model (\cref{sec:eval-virus}), a key benefit of bigraphs is the potential to quickly try out extensions to the model. For example, we could replace the movement system to have the sink move through a dynamic topology (rather than the sensors moving relative to the sink), or we could experiment with data dependent costs by adding structure to the buffer entity to allow it to record the type of data it received.

The existing model does not make use of linking, showing how we can utilise only the parts of the bigraph theory we require, without sacrificing expressiveness.

\section{Discussion}
\label{sec:discussion}

\subsection{Probabilistic Structure}

Our approach is to     generate probabilistic \emph{transition systems} from a
set of probabilistic reaction rules. An alternative  approach is to keep the rewrite semantics unchanged and instead allow the
bigraphs themselves to be probabilistic.

For example, given a non-ground bigraph, we could initialise it
(probabilistically) with different parameters to simulate probabilistic phenomena
and\/or lack of information. 
This is the approach of    term rewriting systems such as
PMaude~\cite{agha.meseguer.ea_PMaude:2006}, where rewriting can affect which
variables are added in the substitution
(the equivalent would be  the instantiation of    bigraph sites). 
This scenario can be modelled in probabilistic bigraphs as a set of rules, one for each possible right-hand side -- where the set of rules act like a probability distribution.

Another approach, due to Syropoulos~\cite{syropoulos2020fuzzy}, is to define the structure of bigraphs (\ie the place and link graphs) in terms of fuzzy sets~\cite{zadeh1965fuzzy}. An advantage of this formalisation is to allow a succinct representation of families of bigraphs. However, it is still an open question how to define bigraphs dynamics, \ie matching and rewriting, in this setting. This makes it difficult to compare it directly with our approach.

\subsection{Comparing Probabilistic/Stochastic Bigraphs and PRISM models}
\label{sec:prismcomp}
Since we use PRISM for analysis of exported systems from BigraphER,  we compare bigraphs with   the PRISM modelling language. 
The latter is based on the language of reactive modules~\cite{AlurH99b}:    a PRISM model     consists of several modules, each with their own
internal state, which are either updated independently or through  
shared (action) names. We make three observations\footnote{We discuss PRISM here, but the arguments are equally applicable to model checkers such as Storm~\cite{storm}.}.

First, consider creating new entities and arbitrary and dynamic topologies. 
PRISM employs \emph{renaming} for this. For example, 
in the original PRISM model for virus spread our model is based on (\cref{sec:eval-virus}), 
  each node is
represented by a module,   links are (shared) actions.  An infected node
attacks by  offering
the action representing a link to another node -- allowing
the state of the attacked node to change.
 New nodes are created   through module  renaming, which  creates a copy of the
module with new state/action names.
But PRISM modules have a fixed \emph{structure}, so, for example, a
module with 2 links (representing a corner node in Fig.~\ref {fig:virus_spread_simple_initial}) cannot be reused for a module with 3 links (an internal node in Fig.~\ref {fig:virus_spread_simple_initial}). Similarly,  
varying the firewall types   requires a new module   for each type of
firewall.
For this reason, creating arbitrary and dynamic topologies in PRISM is difficult and error prone.  On the other  hand, bigraphs allow arbitrary topologies to be  created easily by simply nesting
additional links, and  the graphical notation supports  intuitive checking. This
flexibility also allows    nodes to be added/removed dynamically as
the system progresses over time. 

Second, consider \emph{exact} probabilities.  PRISM   allows these   
to be specified explicitly for \emph{module}  transitions, \eg the probability for a particular 
node to block an infection.
However, in probabilistic bigraphs, probabilities are assigned  to pairs of states of the
entire system, which requires the additional normalisation procedure to account for all
possible (global) transitions. 
PRISM also includes a normalisation procedure to allow multiple modules to be used in parallel, \ie it combines states into a tuple and normalises probabilities~\cite{prism_manual_normalisation}.

In probabilistic bigraphs, as weights are defined as any positive real number, a modeller can use weights of the form $w_1 = p, 0 \le p \le 1$ and $w_2 = 1 - p$ if they prefer. However, we do not enforce transition of this form apply only to a single ``modular'' entity, \ie it is possible to define a single probabilistic transition where multiple entities change state.

Finally, PRISM models are often more efficient to analyse than those constructed from
explicit transition systems, due to specialised  symbolic analysis. It remains an open problem
if similar techniques can be applied  to bigraph models, especially those that  reduce the state space.

\subsection{Future Work}
\label{sec:futurework}

\paragraph{Additional Probabilistic Process Types}
We currently support DTMC, CTMC, and MDP models, however many other 
probabilistic models are used in practice. For example MDPs have been
extended in multiple ways, \eg partially observable
MDPs (POMDPs)~\cite{astrom1965optimal}, and these are candidates for further BRS extensions.

In POMDPs, agents cannot directly observe the current state
and are instead assigned a set of observations/beliefs allowing decision making
in uncertain environments. One possibility for BRSs is to fix the
\emph{actual} and \emph{observed} states into two (bigraph) regions, where cross-region
links connect the observations to the states they observe. Unfortunately while
PRISM allows POMDPs to be specified, there is currently no way to import partially
observable models.

Another extension of MDPs is Markov automata (MA)~\cite{5571733,DENG2013139}, combining probabilistic branching, non-determinism, \emph{and} exponentially distributed delays. In a bigraph model, the probabilistic states of a MA would be treated as the states of an ABRS, while the Markovian states would behave like the states of a SBRS. However, rewriting for \emph{hybrid states}, \ie states where both non-deterministic choice over probability distributions and a distribution over states are available, requires a new definition.

\paragraph{Bisimulations}

Exploring the RPO framework in the presence of probabilistic rules likewise
remains an open problem. It is likely RPOs exist for probabilistic rules,
however these will have to also account for the normalised probabilities of
specific matches (to ensure the contexts are equal).

\section{Conclusion}
\label{sec:conc}

Bigraphical reactive systems (BRSs) have proved invaluable for modelling a wide
range of systems, both virtual and physical, but are limited in the types of
system they can represent. Real-world systems are often probabilistic,
stochastic, and feature non-deterministic decisions, and these do not fit within
standard BRSs.

We have shown how, by assigning \emph{weights} to standard BRS reaction rules we
can model probabilistic systems (probabilistic BRSs -- PBRSs), and extend this
to support systems that make explicit decision (action) choices (action BRSs --
ABRSs).

We have implemented both PBRS and ABRS in BigraphER, an open-source toolkit for
working with bigraphs. To   support stochastic systems we also implement
\emph{stochastic bigraphs} as defined by Krivine
\etal~\cite{DBLP:journals/entcs/KrivineMT08}. We show the new extensions are
practical through a set of case study models: virus spread in computer networks,
membrane budding in biological systems, and data harvesting in wireless sensor
networks.

In conclusion, we have successfully extended the capabilities of  BRSs and the toolkit BigraphER to model a wider range of systems, whilst preserving the   
high level nature and  flexibility of the bigraph formalism.

\section*{Acknowledgement}
This work is supported by the Engineering and Physical Sciences Research
Council, under grant EP/N007565/1 S4: Science of Sensor Systems Software, and by PETRAS SRF grant MAGIC (EP/S035362/1). 
We thank Paulius Dilkas for his work on an early implementation of these ideas.

\Urlmuskip=0mu plus 1mu
\bibliographystyle{alpha}
\bibliography{paper}

\end{document}